\documentclass[11pt, a4paper]{article}

\usepackage{jheparxiv,integrable,mathtools,stmaryrd}
\usepackage[thinlines]{easytable}
\usepackage{tikz-cd}

\usepackage[compat=1.1.0]{tikz-feynman}
\usetikzlibrary{arrows,positioning,arrows.meta}
\tikzfeynmanset{ with arrow/.style = {
   decoration={
     markings,
     mark=at position 0.5
          with {\arrow[xshift=1mm]{stealth'[width=2mm,length=3mm]}}
     },
   postaction=decorate}
}

\newcommand{\tbar}{{\bar{t}{}}}
\newcommand\scalemath[2]{\scalebox{#1}{\mbox{\ensuremath{\displaystyle #2}}}}

\subheader{\hfill \texttt{DAMTP-2019}}

\title{Gauge Theory and Boundary Integrability II:\\ Elliptic and Trigonometric Cases}

\author{Roland Bittleston and David Skinner}

\affiliation{Department of Applied Mathematics \& Theoretical Physics \\
	University of Cambridge \\
	Wilberforce Road \\
	Cambridge CB3 0WA, United Kingdom}

\emailAdd{[r.bittleston, d.b.skinner]@damtp.cam.ac.uk}

\abstract{We consider the mixed topological--holomorphic Chern--Simons theory introduced by Costello, Yamazaki \& Witten on a $\Z_2$ orbifold. We use this to construct semi-classical solutions of the boundary Yang--Baxter equation in the elliptic and trigonometric cases.  A novel feature of the trigonometric case is that the $\Z_2$ action lifts to the gauge bundle in a $z$-dependent way. We construct several examples of $K$-matrices, and check they agree with cases appearing in the literature.}

\begin{document}

\maketitle

%\newpage

\section{Introduction}
In~\cite{Bittleston:2019gkq} rational solutions of the boundary Yang-Baxter equation were generated as the vacuum expectation values of Wilson lines in a mixed topological-holomorphic analogue of Chern-Simons theory on a $\Z_2$-orbifold. This extended the link between gauge theory and quantum integrability developed by Costello, Witten, and Yamazaki in the papers~\cite{Costello:2013zra,Costello:2013sla,Costello:2017dso,Costello:2018gyb} to integrable models with boundary. In this paper we expand this construction to elliptic and trigonometric solutions of the boundary Yang-Baxter equation.

We begin by reviewing the CWY approach to 2d quantum integrable lattice models. We shall be brief, and refer the reader to~\cite{Costello:2013zra,Costello:2017dso} for more detailed discussions of the theory

\subsection{CWY Theory}
\label{sec:CWYTheory}

CWY theory is defined on a four-manifold $M=\Sigma \times C$, where $\Sigma$ is some 2-dimensional real manifold and $C$ is a Riemann surface admitting a closed, holomorphic 1-form $\omega$. The condition that $\omega$ be closed, holomorphic, and nowhere vanishing on $C$ is very restrictive, forcing $(C,\omega)$ to be biholomorphic to $(\mathbb{C},\d z)$, $(\mathbb{C}^*,\d u/ u=\d z)$, or $(E_\tau,\d z)$. These three choices generate rational, trigonometric, and elliptic quasi-classical solutions to the YBE, respectively. In this paper will concentrate on the trigonometric and elliptic cases. We will often use real coordinates ${\bf r} = (x,y)$ on $\Sigma$, and complex coordinates $(z,\bar z)$ on $C$. We denote these coordinates collectively by $w\in M$.  

The action of CWY theory is
\begin{equation}
\label{CSAction}
S_M[A] = \frac{1}{2\pi}\int_{M}\omega \wedge 
\Tr\bigg(A\wedge \diff A + \frac{2}{3}A\wedge A\wedge A\bigg) 
= \frac{1}{2\pi}\int_{M}\omega\wedge\text{CS}(A)\,,
\end{equation}
where 
\[
	A(w) = A_x(w) \d x + A_y(w) \d y + A_{\bar z}(w)\d\bar z
\]
is a partial connection on a $G$-bundle over $M$. Note that $A$ is only a partial connection since it has no $A_z$ component. The theory is topological in $\Sigma$ and holomorphic in $C$, {\it i.e.} it is invariant under diffeomorphisms of $M$ which are independent of $C$.  We can only study this theory perturbatively,  for example because the periods of $\omega/\hbar$ are not naturally quantized,

%% In fact it is possible to write down a more general class of line operators by viewing $A$ as taking values in an enlarged Lie algebra. To do this we expand the gauge field as a formal power series in a parameter $u$
%%\[ A^{\fg[[u]]}(w) = \sum_{n=0}^\infty\frac{u^nt_a}{n!}(\partial_z^n A^a)(w)\,,\]
%%where \smash{$\{t_a\}_{a=1,\dots,\dim\fg}$} is a basis of $\fg$. We can view $A^{\fg[[u]]}$ as 1-from taking values in the infinite dimensional Lie algebra of formal power series in $u$ with coefficients in $\mathfrak{g}$, denoted $\fg[[u]] = \prod_{n=0}^\infty(\fg\otimes u^n)$. This infinite dimensional Lie has a (topological) basis \smash{$\{u^nt_a\}_{a=1,\dots,\dim\fg,\,n\geq0}$}. We can then write down generalised Wilson lines, which are still supported at fixed points in $C$, but which are associated to a representation $V$ of $\fg[[u]]$
%%\[ \mathcal{W}_V[\gamma,z_0] = P\exp\left(\int_{\gamma\times\{z_0\}}A^{\fg[[u]]}_V\right) = P\exp\left(\int_{\gamma\times\{z_0\}}\sum_{n=0}^\infty\frac{t_{a,V}^{(n)}}{n!}(\partial_z^n A^a)(w)\right)\,.\]
%%In the quantum theory Wilson lines suffer from anomaly at order $\hbar^2$. To cancel this anomaly the $t_{a,V}^{(n)}$ must be chosen to obey the defining relations of the Yangian, a deformation of the universal enveloping algebra $U(\mathfrak{g}[[u]])$ at order $\hbar^2$.

The simplest observables in CWY theory are Wilson lines 
\[ 
	\mathcal{W}_V[\gamma,z_0] = P\exp\left(\int_{\gamma\times\{z_0\}} A_V\right)\,,
\]
supported on a straight line $\gamma\subset\Sigma$ and at point $z_0\in C$. The vacuum expectation value of the following configuration of crossing Wilson lines
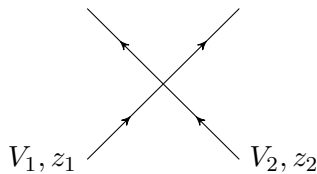
\begin{figure}[h!]
	\centering
	\begin{tikzpicture}[baseline]
	\begin{feynman}
	\vertex(o);
	\vertex[right=1cm of o](e);
	\vertex[left=1cm of o](w);
	\vertex[right=1cm of o](oe);
	\vertex[left=1cm of o](ow);
	\vertex[above=1cm of e](ne);
	\vertex[below=1cm of e](se);
	\vertex[above=1cm of w](nw);
	\vertex[below=1cm of w](sw);
	\vertex[below=1cm of oe](ose);
	\vertex[below=1cm of ow](osw);
	\vertex[right=0cm of se]{$V_2,z_2$};
	\vertex[left=0cm of sw]{$V_1,z_1$};
	\diagram*{(o) -- [fermion] (ne), (se) -- [fermion] (o), (sw) -- [fermion] (o), (o) -- [fermion] (nw)};
	\end{feynman}
	\end{tikzpicture}
\end{figure}\\
generates an $R$-matrix
\[
	R_{12}(z_1-z_2) : V_1\otimes V_2\to V_1\otimes V_2\,.
\] 
IR freedom of the theory after gauge fixing together with diffeomorphism invariance in $\Sigma$ ensures that the $R$-matrix is local to the point of crossing, so that vevs of configurations in which several Wilson lines cross reduce to the product of $R$-matrices. The $R$-matrices automatically obey  the Yang-Baxter equation
\begin{figure}[h!]
	\centering
	\begin{tikzpicture}[baseline]
	\begin{feynman}
	\vertex (o){$=$};
	\vertex[left=3cm of o] (l);
	\vertex[above=1.3cm of l] (lbf);
	\vertex[below=1.3cm of l] (lbi);
	\vertex[left=0.75cm of l](ll);
	\vertex[above=0.86cm of ll](lcf);
	\vertex[below=0.86cm of ll](lai);
	\vertex[right=1.5cm of l](lr);
	\vertex[above=0.42cm of lr](laf);
	\vertex[below=0.42cm of lr](lci);
	\vertex[right=3cm of o] (r);
	\vertex[above=1.3cm of r] (rbf);
	\vertex[below=1.3cm of r] (rbi);
	\vertex[right=0.75cm of r](rr);
	\vertex[above=0.86cm of rr](raf);
	\vertex[below=0.86cm of rr](rci);
	\vertex[left=1.5cm of r](rl);
	\vertex[above=0.42cm of rl](rcf);
	\vertex[below=0.42cm of rl](rai);
	\diagram*{{[edges=fermion] (lai) --  (laf), (lbi) --  (lbf), (lci) --  (lcf), (rai) --  (raf), (rbi) --  (rbf), (rci) --  (rcf)},};
	\vertex[left=0.1cm of lai]{$V_1,z_1$};
	\vertex[below=0.1cm of lbi]{$V_2,z_2$};
	\vertex[below=0.1cm of lci]{$V_3,z_3$};
	\vertex[below=0.1cm of rai]{$V_1,z_1$};
	\vertex[below=0.1cm of rbi]{$V_2,z_2$};
	\vertex[right=0.1cm of rci]{$V_3,z_3$};
	\end{feynman}
	\end{tikzpicture}
\end{figure}\\
since no singularities are encountered in any Feynman diagram when moving between the two configuration. Algebraically, this equation is
\[
R_{12}(z_1-z_2)R_{13}(z_1-z_3)R_{23}(z_2-z_3) 
= R_{23}(z_2-z_3)R_{13}(z_1-z_3)R_{12}(z_1-z_2)
\]
as is well-known.

Since the mixed topological--holomorphic Chern--Simons theory is defined only perturbatively, the $R$-matrices it generates inevitably take the form
\begin{equation}
\label{eq:qcrm} 
	R_\hbar(z) = {\bf 1}_{V\otimes V'} + \hbar\,r_{V\otimes V'}(z) 
+ \mathcal{O}(\hbar^2)\,.
\end{equation}
$R$-matrices admitting\footnote{Note that there exist solutions to the YBE which are not quasi-classical.} such a formal expansion in a parameter $\hbar$ are called quasi-classical, and $r(z)$ is known as the classical $r$-matrix. This classical $r$-matrix is assumed to take values in $\fg\otimes\fg$ for $\fg$ a finite dimensional, complex, reductive Lie algebra, acting in a product $V\otimes V'$ of representations of $\fg$. Expanding the YBE to second order in $\hbar$ shows that the classical $r$-matrix obeys the classical Yang-Baxter equation,
\[
[r_{13}(z_1-z_3),r_{23}(z_2-z_3)] + [r_{12}(z_1-z_2),r_{13}(z_1-z_3)] 
+ [r_{12}(z_1-z_2),r_{23}(z_2-z_3)] = 0\,.
\]
Solutions of the classical Yang-Baxter equations were essentially classified by Belavin \& Drinfeld~\cite{Belavin:1982cybe}, under the assumption that the $r$-matrix is non-degenerate. The solutions can be separated into three families, distinguished by whether the classical $r$-matrix can be written in terms of rational, trigonometric, or elliptic functions. In this work we will concentrate on the trigonometric and elliptic cases.

The easiest way to compute the semi-classical contribution to the $R$-matrix in CWY theory is to work in `holomorphic gauge'
\[ 
	A_{\bar z} = 0\,.
\]
Much like axial gauge in ordinary Chern--Simons theory, this is not a good gauge for performing loop computations, but it makes finding the semi-classical contribution to the $R$-matrix essentially trivial~\cite{Costello:2019tri}. In holomorphic gauge, the propagator is
\[ 
	P_{xy}(w;w') = \delta^{(2)}_{\Sigma}({\bf r} - {\bf r}')\,r(z-z')\,,
\]
where $r(z-z')$ is defined by 
\[ 
	\partial_{\bar z}(r(z-z')) = c\,\delta^{(2)}(z-z') 
\]
and $c\in\fg\otimes\fg$ is the inverse of the $\fg$-invariant bilinear used to define the theory. Computing the order $\hbar$ contribution to the quasi-classical $R$-matrix using this propagator gives 
\[ 
	R_{12}(z_1-z_2) = {\bf 1}_{V_1\otimes V_2} + \hbar\, r_{V_1\otimes V_2}(z_1-z_2) + \mathcal{O}(\hbar^2)\,,\]
and we can see that $r(z)$ appearing in the propagator is in fact the corresponding classical $r$-matrix.

Abusing notation by writing ${\rm Ad}\,P\to C$ for the adjoint bundle of the gauge theory an arbitrary point in $\Sigma$, $r(z-z')$ should be interpreted as a meromorphic section of $\text{Ad}(P)\boxtimes\text{Ad}(P)\to C \times C$. In the above formula $\partial_{\bar z}$ refers to the partial connection on ${\rm Ad}\,P$ determined by the vacuum around which we are expanding, lifted to act on the first factor of $\text{Ad}(P)\boxtimes\text{Ad}(P)$. In this way $r(z)$ depends on the choice of adjoint bundle ${\rm Ad} P$, and our choice of vacuum. For example, when $C=\C$, ${\rm Ad}\,P$ is the trivial bundle on $M$ whose sections are required to tend to 0 at infinity. In this case we have
\[ r(z) = \frac{c}{z}\,,\]
which is the rational classical $r$-matrix. 

In~\cite{Costello:2018gyb} it was demonstrated that, for all three choices of $C$, the classical $r$-matrix, together with the formal properties satisfied by the full quasi-classical $R$-matrix, is enough to fix the quasi-classical $R$-matrix to all orders in $\hbar$ up to ambiguities which have a natural interpretation in terms of the parameters defining the theory.

\subsection{The boundary Yang-Baxter equation}
\label{sec:bYBE}

In this paper we will be concerned with integrable lattice models with boundaries, and with boundary conditions preserving their bulk integrability. Investigations of such boundary conditions date back to work of Skylanin~\cite{Sklyanin:1988yz} and Olshanski~\cite{Olshanskii:1992tw,Molev:1994rs}, and have been extensively studied since. Boundary conditions on a spin chain are encoded in a $K$-matrix 
\[
	K(z):V\otimes W\to V'\otimes W
\] 
which again depends meromorphically on the spectral parameter $z$. Here we interpret $V$ and $V'$ as the state spaces of an incoming and outgoing excitation reflecting off the boundary, and $W$ as a space of boundary states.  The boundary conditions preserve integrability if the $K$-matrix obeys the boundary Yang-Baxter equation
\begin{equation}
\label{bYBE}
	R_{12}(z_1-z_2)K_{13}(z_1)R_{21}(z_1+z_2)K_{23}(z_2) 
	= K_{23}(z_2)R_{12}(z_1+z_2)K_{13}(z_1)R_{21}(z_1-z_2)\,,
\end{equation}
which may be viewed pictorially as\\

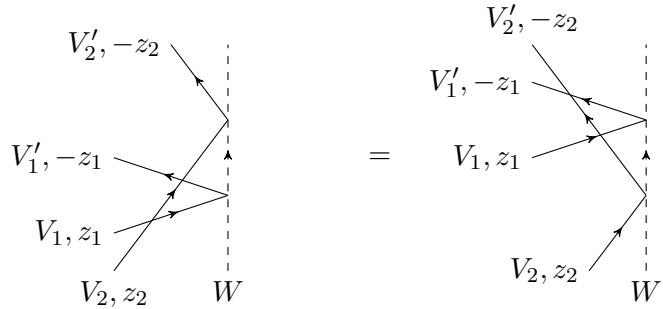
\vspace{-0.5cm}
\begin{figure}[h!]
	\centering
	\begin{tikzpicture}[baseline]
	\begin{feynman}
	\vertex (o){$=$};
	\vertex[left=2cm of o](l);
	\vertex[above=0.5cm of l](lb);
	\vertex[below=0.5cm of l](la);
	\vertex[above=1.5cm of l](ln);
	\vertex[below=1.5cm of l](ls);
	\vertex[left=1.5cm of ls](lsw);
	\vertex[left=1.5cm of l](lw);
	\vertex[below=1cm of lw](lai);
	\vertex[left=0.75cm of ln](lbf);
	\vertex[right=3.5cm of o](r);
	\vertex[above=0.5cm of r](ra);
	\vertex[below=0.5cm of r](rb);
	\vertex[above=1.5cm of r](rn);
	\vertex[below=1.5cm of r](rs);
	\vertex[left=1.5cm of rn](rnw);
	\vertex[left=1.5cm of r](rw);
	\vertex[above=1cm of rw](raf);
	\vertex[left=0.75cm of rs](rbi);
	\diagram*{{[edges=fermion] (lai) --  (la), (la) --  (lw), (lsw) --  (lb), (lb) --  (lbf), (rw) -- (ra), (ra) -- (raf), (rbi) -- (rb), (rb) -- (rnw)},{[edges=charged scalar] (ls) -- (ln), (rs)--(rn)},};
	\vertex[below=0cm of lsw]{$V_2,z_2$};
	\vertex[left=0cm of lai]{$V_1,z_1$};
	\vertex[left=0cm of lw]{$V_1',-z_1$};
	\vertex[left=0cm of lbf]{$V_2',-z_2$};
	\vertex[left=0cm of rbi]{$V_2,z_2$};
	\vertex[left=0cm of rw]{$V_1,z_1$};
	\vertex[left=0cm of raf]{$V_1',-z_1$};
	\vertex[above=0cm of rnw]{$V_2',-z_2$};
	\vertex[below=0cm of ls]{$W$};
	\vertex[below=0cm of rs]{$W$};
	\end{feynman}
	\end{tikzpicture}
\end{figure}
\noindent Note that when a line reflects of the boundary its spectral parameter changes sign.  We will be primarily concerned with the case $V_i\cong V'_i$.

In~\cite{Bittleston:2019gkq} we showed that solutions of the boundary Yang-Baxter equation may be obtained by placing CWY theory on the orbifold
\smash{$\widetilde{M}=M/\Z_2=(\Sigma\times C)/\Z_2$}. In defining this orbifold,  we assume that the 2-dimensional smooth manifold $\Sigma$ is the double of some manifold with boundary $\overline\Sigma$. This means that $\Sigma$ can be formed by gluing together two copies of $\overline\Sigma$ along their boundary. $\Sigma$ hence admits a natural reflection that swaps the two copies of $\Sigma$ whilst fixing their common boundary.\footnote{To avoid a global anomaly we also require that $\overline\Sigma$ be framed in the sense that $\Sigma$ admits a nowhere vanishing vector field that is equivariant under this reflection. Such $\overline\Sigma$ include the half-plane, half-cylinder, and annulus.} For convenience we choose our coordinates $(x,y)$ on $\Sigma$ so that the reflection acts as $(x,y)\mapsto (-x,y)$. Then the generator $Z$ of the $\Z_2$ symmetry on $\widetilde M$ is
\[ 
	Z : (x,y,z,\bar z)\mapsto (-x,y,-z,-\bar z)\,.
\]
Note that the fixed points of this action, and hence the singular points on the orbifold, are lines in $\Sigma$ fixed by $Z$ supported at fixed points of $z\mapsto-z$ on $C$. For $C=\C$ there is one such fixed point, for $C=\C^*$ there are two fixed points, and for $C=E_\tau$ there are four fixed points.

To lift the action of $Z$ to the adjoint bundle of the gauge theory, in~\cite{Bittleston:2019gkq} we simultaneously acted of the fibres of this bundle with an automorphism $\sigma\in{\rm Aut}(\fg)$. The condition that this defines an action of $\Z_2$ restricts the automorphism to be involutive. In full the action of $Z$ on the gauge field is
\[ Z:A\mapsto\sigma(Z^*\!A)\]
To do gauge theory on the orbifold we only integrate over field configurations on $\Sigma\times C$ which are fixed by this $\Z_2$ action. We use the action $S[A]/2$, which is a consistent as long as
\[ S[\sigma(Z^*\!A)] = S[A]\,.\]
This holds if $\sigma$ preserves the $\fg$-invariant bilinear used in defining the action. In section~\ref{sec:trigk2} we will generalise this picture somewhat.

The condition $A=\sigma(Z^*\!A)$ in particular implies that the gauge field along the orbifold fixed lines takes values in a subalgebra  $\mathfrak{p}\subset\fg$ that is fixed by $\sigma$. Thus, along such lines we can introduce a new family of line operators that (in the simplest case\footnote{In the rational case, we showed in~~\cite{Bittleston:2019gkq} that boundary Wilson lines really live in representations of the twisted Yangian $\mathcal{B}(\mathfrak{p},\fg)$.}) live in representations of this subalgebra. We refer to these operators as `boundary Wilson lines'. Observables in the theory are configurations of bulk Wilson lines in $M$, together with boundary Wilson lines inserted along the lines of singular points. This configuration of bulk and boundary Wilson lines must be chosen so that no operators coincide in $\widetilde M$.

CWY theory on an orbifold generates $K$-matrices via the vacuum expectation values of the following configurations of Wilson lines:
\newpage
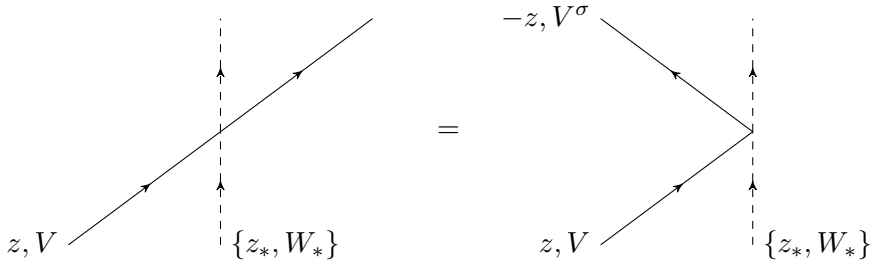
\begin{figure}[th]
	\centering
	\begin{tikzpicture}[baseline]
	\begin{feynman}
	\vertex(c){$=$};
	\vertex[left=3cm of c](lo);
	\vertex[left=2cm of lo](lw);
	\vertex[right=2cm of lo](le);
	\vertex[below=1.5cm of lw](lsw);
	\vertex[above=1.5cm of le](lne);
	\vertex[below=1.5cm of lo](ls);
	\vertex[above=1.5cm of lo](ln);
	\vertex[left=0cm of lsw]{$z,V$};
	\vertex[right=0cm of ls]{$\{z_*,W_*\}$};
	\diagram*{{[edges=fermion]  (lsw) -- (lo), (lo) -- (lne)}, {[edges = charged scalar] (ls) -- (lo), (lo) -- (ln)},};
	\vertex[right=4cm of c](re);
	\vertex[left=2cm of re](rw);
	\vertex[above=1.5cm of rw](rnw);
	\vertex[below=1.5cm of rw](rsw);
	\vertex[above=1.5cm of re](rne);
	\vertex[below=1.5cm of re](rse);
	\vertex[left=0cm of rsw]{$z,V$};
	\vertex[left=0cm of rnw]{$-z,V^\sigma$};
	\vertex[right=0cm of rse]{$\{z_*,W_*\}$};
	\diagram*{{[edges=fermion]  (rsw) -- (re), (re) -- (rnw)}, {[edges = charged scalar] (rse) -- (re), (re) -- (rne)},};
	\end{feynman}
	\end{tikzpicture}
\end{figure}

\noindent The picture on the left represents the Wilson line configuration on $M$, the double cover of the orbifold. This picture is often useful for calculation. The second picture is the more familiar depiction of the $K$-matrix as describing the reflection of an excitation off a boundary. To obtain it, we have used the $\Z_2$ action to write all the observables in the region\footnote{If we so wish, we can then view the theory as being defined on the manifold with boundary $M_{\rm L}$, albeit with non-local boundary conditions imposed on $A$ at $x=0$.} $M_{\rm L}=\Sigma_{x\leq0}\times C$. Note, however, that the choice $M_{\rm L}$ is artificial and is not determined canonically by the orbifold structure. In each case, $\{z_*,W_*\}$ indicates the presence of a set of boundary Wilson lines, one at each singular point in $C$, in representations $W_*$ of $\mathfrak{p}$. (Thus $\{z_*,W_*\}$ indicates one, two or four such boundary lines in the rational, trigonometric and elliptic cases, respectively.)

\medskip

Since we work perturbatively, as for the bulk $R$-matrix,  the $K$-matrices we generate inevitably admit an expansion in $\hbar$.  However, unlike the quasi-classical $R$-matrices, the leading ($\hbar=0$) term will not typically be the identity. This is because, in order to obtain the more familiar (second) picture, we have acted on $V$ with the bundle automorphism $\sigma$. In particular, suppose $\sigma$ is an inner automorphism so that $V^\sigma\cong V$. Then, with a single boundary Wilson line, the usual $K$-matrix would be viewed as a map 
\[
	K(z) : V\otimes W\to V\otimes W\,,
\]
with the two copies of $V$ canonically identified. However, in our case the identification is non-trivial: inner automorphisms correspond to conjugation by some $\tau$ (for $\sigma$ to be an involution we require $\tau^2\in Z(G)$). Then $V^\sigma$ identified with $\tau_V \,V$ where $\tau_V$ is $\tau$ acting in the $V$ representation. Thus, the vacuum expectation of the configuration above computes \smash{$K(z)\,\tau^{-1}_V$} rather than the usual $K$-matrix itself. In other words, the semi-classical $K$-matrix generated by CWY theory on an orbifold takes the form
\[
	K_\hbar(z-z_*) = {\rm Feyn}(z-z_*) \, \big(\tau_V\otimes {\bf 1}_{W}) 
	= \tau_V\otimes {\bf 1}_{W} + \hbar\, k_{V\otimes W}(z-z_*) 
	+ \mathcal{O}(\hbar^2)\,,
\]
where ${\rm Feyn}(z-z_*)$ is the sum of all Feynman diagrams contributing to the expectation value. As in the bulk, these $K$-matrices inevitably obey the boundary Yang-Baxter equation~\eqref{bYBE}.

The semi-classical contributions to the vacuum expectation of the above diagrams are again easy to evaluate in holomorphic gauge. We represent the configuration of Wilson lines on the orbifold via the picture on the left above. From the method of images, a propagator which respects the $\Z_2$ orbifold action is given by
\[ 
	P_{xy}(w;w') = \delta^{(2)}_\Sigma\!({\bf r} - {\bf r'})\,r(z-z') 
	+ \delta^{(2)}_\Sigma\!\big({\bf r} - Z({\bf r'})\big)\,\sigma_2(r(z+z')) \,,
\]
where $r(z)$ is the bulk classical $r$-matrix and $\sigma_2$ denotes our automorphism acting on the second factor.  In the presence of $n$ boundary Wilson lines there are $1+n$ diagrams which contribute. They are the bulk self-interaction and the $n$ bulk-boundary interactions\footnote{We neglect self-interactions on the bulk line which do not arise as a result of the orbifold structure and simply alter its normalization. Similarly we neglect self-interactions of the boundary lines.}.

The bulk self-interaction arises from the contribution to the following diagram of the image part of the propagator
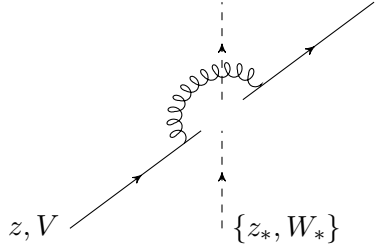
\begin{figure}[th]
	\centering
	\begin{tikzpicture}[baseline]
	\begin{feynman}
	\vertex (lo);
	\vertex[left=2cm of o](w);
	\vertex[right=2cm of o](e);
	\vertex[left=0.5cm of o](lw);
	\vertex[right=0.5cm of o](le);
	\vertex[below=1.5cm of w](sw);
	\vertex[above=1.5cm of e](ne);
	\vertex[below=0.375cm of lw](lsw);
	\vertex[above=0.375cm of le](lne);
	\vertex[below=1.5cm of o](s);
	\vertex[above=1.5cm of o](n);
	\vertex[left=0cm of sw]{$z,V$};
	\vertex[right=0cm of s]{$\{z_*,W_*\}$};
	\diagram*{{[edges=fermion]  (sw) -- (o), (o) -- (ne)}, {[edges = charged scalar] (s) -- (o), (o) -- (n)},{[edges = gluon] (lsw) -- [half left] (lne)}};
	\end{feynman}
	\end{tikzpicture}
\end{figure}\\
In terms of the basis $\{t_a\}_{a=1}^{\dim\fg}$ of $\fg$, the contribution of this diagram is
\[ \begin{aligned}
&\left(\sum_{a,b=1}^{\dim\fg} t_a \,\sigma(t_b)\right)\Bigg\rvert_V \ r^{ab}(2z) \int_{s\leq t}\diff s \diff t\, 2\sin\theta\cos\theta\,\delta((t+s)\cos\theta)\,\delta((t-s)\sin\theta)  \\
&=\frac{1}{2}\left(\sum_{a,b=1}^{\dim\fg} t_a \,\sigma(t_b)\right)\Bigg\rvert_Vr^{ab}(2z) =  
\frac{1}{2}\,{\rm gl}\big(\sigma_2(r(2z)\big)|_V\,,
\end{aligned} \]
where the linear map \smash{${\rm gl}:U(\fg)\otimes U(\fg)\to U(\fg)$} acts on $X\otimes Y$ by gluing the two factors of the tensor product together in order to give $XY$. (Note that ${\rm gl}$ is {\it not} an isomorphism of associative algebras.)

The remaining diagrams describe bulk-boundary interactions, and there as many of these diagrams as there are boundary Wilson lines. They are all represented by the diagram below.
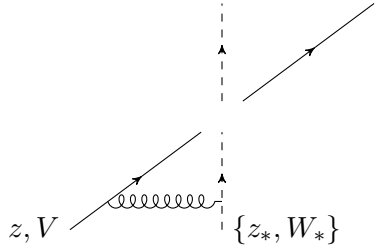
\begin{figure}[th]
	\centering
	\begin{tikzpicture}[baseline]
	\begin{feynman}
	\vertex (lo);
	\vertex[left=2cm of o](w);
	\vertex[right=2cm of o](e);
	\vertex[below=1.5cm of w](sw);
	\vertex[above=1.5cm of e](ne);
	\vertex[below=1.5cm of o](s);
	\vertex[above=1.5cm of o](n);
	\vertex[below=1.125 of o](ls);
	\vertex[left=1.5cm of ls](lsw);
	\vertex[left=0cm of sw]{$z,V$};
	\vertex[right=0cm of s]{$\{z_*,W_*\}$};
	\diagram*{{[edges=fermion]  (sw) -- (o), (o) -- (ne)}, {[edges = charged scalar] (s) -- (o), (o) -- (n)},{[edges = gluon] (lsw) -- (ls)}};
	\end{feynman}
	\end{tikzpicture}
\end{figure}\\
Inserting the boundary Wilson lines at singular points $z_*$ ensures that, when coupled to by a propagator, both the bulk and mirror part of the propagator contribute identically. This means that we can effectively replace the propagator by \smash{$2\,\delta^{(2)}_\Sigma\!({\bf r} - {\bf r}')\,r(z-z')$} so that the contribution of the above diagram is just $2r(z-z_*)|_{V\otimes W_*}$. It is then clear that full semi-classical limit of the $K$-matrix is given by
\be 
\label{eq:lmx} 
	K(z)\tau_V^{-1} = {\bf 1}_{V\otimes W} + \frac{\hbar}{2}{\rm gl}\big(\sigma_2(r(2z)\big)|_V + 2\hbar\sum_{z_*}r(z-z_*)|_{V\otimes W_*} + \mathcal{O}(\hbar^2)\,,
\ee
where $\sigma = {\rm conj_\tau}$. (This formula also applies for $\sigma$ outer, but we must work a little harder to interpret it correctly.) We will sometimes refer to the semi-classical $K$-matrix $k(z)$ as $\ell(z) \tau_V$, so that
\[
	\ell(z) = \ell_0(z) + \delta\ell(z)
	= \frac{1}{2}{\rm gl}\big(\sigma_2(r(2z)\big)|_V + 2 \sum_{z_*}r(z-z_*)|_{V\otimes W_*}
\]
with $\ell_0(z)$ the order $\hbar$ self-interaction term and $\delta\ell(z)$ the corresponding bulk-boundary interaction.

\section{Elliptic Solutions}
\label{sec:elliptic}
In this section we will discuss how to generate elliptic $K$-matrices using gauge theory. We start by reviewing the method for generating elliptic $R$-matrices described in~\cite{Costello:2017dso}.

\subsection{Elliptic solutions of the Yang-Baxter equation}
\label{sec:ellR}

The gauge invariant data determining a classical solution of CWY theory is a stable holomorphic $G$-bundle over $C$. For the vacuum to be isolated, the tangent space to the moduli space of such bundles at the vacuum must be trivial. When $C=E_\tau$, the topological class of such a bundle is determined by an element of $\pi_1(G)$. The stability condition amounts to requiring that the Lie algebra of the automorphism group of the holomorphic $G$-bundle be trivial, or equivalently that the automorphism group be discrete. This constrains the gauge group to be $G=\text{PSL}_n(\C)\cong\pi_0(\text{Aut}(\mathfrak{sl}_n\C))$ for $n\geq2$, and the topological class of the $\text{PSL}_n(\C)$-bundle to be a generator $\zeta\in\Z_n\cong\pi_1(\text{PSL}_n(\C))$. 

To construct the bundle explicitly we introduce the matrices $A, B\in \text{SL}_n\C$ obeying
\[ 
ABA^{-1}B^{-1} = \varepsilon\,,
\]
where $\varepsilon=\exp(2\pi i/n)$ is an $n^\text{th}$ root of unity\footnote{What we're really doing here is choosing a conjugacy class in the set of pairs of commuting elements in $\text{PSL}_n(\C)$. This is precisely the data required to define a flat ${\rm PSL}_n(\C)$ bundle over an elliptic curve.}. Up to conjugation and scalar multiplication, we can without loss of generality take the components of $A$ and $B$ to be
\[ 
A^\alpha_{~\beta} = \varepsilon^\alpha\delta_{\alpha,\beta}\,,\qquad B^\alpha_{~\beta} = \delta_{\alpha,\beta+1}\,.
\]
Viewed as elements of $\text{PSL}_n(\C)$, $A$ and $B$ commute, and hence can be used to define a flat $G$-bundle over $E_\tau$ with monodromies $A^\zeta$ and $B$ around the two cycles. Forgetting the $\partial_z + A_z$ component of the covariant derivative, we get a holomorphic $G$-bundle over $\Sigma$ whose automorphism group is $\Z_n^2$. This is the vacuum we will expand around.

As a holomorphic bundle over $E_\tau$ the adjoint bundle of the gauge theory is given by
\[ 
	\text{Ad}(P) = \bigoplus_{(i,j)\in\mathcal{I}_n}\mathcal{L}_{i,j}\otimes t_{i,j}\,,
	\]
where $\mathcal{I}_n=\Z_n^2\setminus\{(0,0)\}$ while $t_{i,j}=B^{\zeta^{-1}i}A^{-j}$. The $\mathcal{L}_{i,j}$ are holomorphic line bundles over the elliptic curve with vanishing Chern class and corresponding to the points $(i + j\tau)/n$ in the Jacobian variety. The set $\{t_{i,j}\}_{(i,j)\in\mathcal{I}_n}$ form a basis of $\mathfrak{sl}_n\C$. This basis obeys
\[ 
	A^\zeta \,t_{i,j}\,A^{-\zeta}  = \varepsilon^i \,t_{i,j}\,,\qquad 
	B\,t_{i,j}\,B^{-1} = \varepsilon^j t_{i,j}\,.
\]
and so diagonalizes the conjugacy action of $A^\zeta$ and $B$.

Now we've fixed the vacuum we can construct the propagator and determine the next-to-leading order contribution to the quasi-classical $R$-matrix. Recall that in holomorphic gauge the relevant component of the propagator is
\[ 
	P_{xy}(w,w') = r(z-z')\,\delta^{(2)}({\bf r} - {\bf r'}) 
\]
where $r(z)$ is the elliptic classical $r$-matrix taking the form
\be \label{eq:ellrmx} r(z) = \frac{1}{n}\sum_{(i,j)\in\mathcal{I}_n}e^{-\zeta^{-1}ij}\,t_{i,j}\otimes t_{-i,-j} \,w_{i,j}(z) \,. \ee
Here the $w_{i,j}(z)$ are given by
\[ w_{i,j}(z) = \sum_{p,q\in\Z_n^2}\frac{\varepsilon^{ip+jq}}{z-p-q\tau}\,.\]
The $w_{i,j}(z)$ obey the quasi-periodicity conditions
\[ 
	w_{i,j}(z+1) = \varepsilon^iw_{i,j}(z)\,,\qquad 
	w_{i,j}(z+\tau) = \varepsilon^jw_{i,j}(z)\,.
\]
and so are meromorphic sections of $\cL_{i,j}$. They are the unique such sections with a simple pole with residue 1 at the origin, as befits the propagator.

\subsection{Elliptic solutions of the boundary Yang-Baxter equation} \label{subsec:ellorb}
Now consider CWY theory on the orbifold ${\widetilde M} = (\Sigma\times E_\tau)/\mathbb{Z}_2$, where we recall that the $\Z_2$ acts as
\[ 
	Z: (x,y,z,\bar{z}) \mapsto (-x,y,-z,-\bar{z})\,.
\]
In lifting this $\Z_2$ action to the adjoint bundle we have the freedom to simultaneously act with the involutive bundle morphism
\[ 
	\sigma: \text{Ad}\,P \to \text{Ad}\,P
\]
preserving the vacuum gauge field $A=0$. It is straightforward to determine that $\sigma$ acts on the fibres as a constant involutive automorphism of $\mathfrak{sl}_n\C$ which inverts the monodromies around the two cycles
\[ 
	\sigma(A) \propto A^{-1}\,,\qquad\sigma(B)\propto B^{-1}\,.
\]
This then fixes $\sigma$ in terms of two parameters $\xi,\eta\in\mathbb{Z}_n$ to be
\begin{equation}
\label{eqn:ellipticsigma}
\sigma(t_{i,j}) = \varepsilon^{i\xi + j\eta}\,t_{-i,-j}\,.
\end{equation}
A derivation of this is included in appendix~\ref{app:EllipticBundleMorphisms}.

\medskip

Now let's consider the behaviour of the gauge field at orbifold singularities. These are the four lines $L_*=\{x=0,\,z=z_*\}$ for $z_*=(a_* + b_*\tau)/2$ with $a_*,b_* \in\{0,1\}$. The quotient $E_\tau/\{z\sim-z\}$ has four conical singularities corresponding to the four fixed points, and embeds in 3-dimensional space as a `pillowcase'.  Pulling back the gauge field to these lines we learn that\footnote{Note that the ambiguity in the choice of map $z_*\mapsto -z_*$ can be absorbed into a different choice of $\sigma$.}
\[ 
	A_y(0,y,z_*,\bar{z}_*) = \sigma(A_y)(0,y,-z_*,-{\bar z}_*) 
	= \sigma_*(A_y)(x,0,z_*,{\bar z}_*)\,,
\]
where we have defined $\sigma_*= \sigma\circ\text{conj}_{A^{-\zeta}}^{a_*}\circ\text{conj}_{B^{-1}}^{b_*}$ and $\text{conj}_{B^{-1}}^{b_*}$ means we conjugate by $B^{-1}$ if $b_*=1$. This shows that the subalgebra of $\mathfrak{sl}_n\C$ in which $A_y$ takes values depends on which of the four singular lines we pull back to. Since we have
\[ 
	\sigma_*(t_{i,j}) = \varepsilon^{i(\xi-a_*)+j(\eta-b_*)}t_{-i,-j}\,,
\]
the effect of $\sigma_*$ is to shift the non-trivial monodromies as 
\[
	\xi\mapsto\xi - a_* = \xi_*\qquad\text{and} \qquad\eta\mapsto\eta - b_* = \eta_*\,
\]
The positive eigenspace of this automorphism, which we shall denote by $\fp_*$ depends on $n$, $\xi_*$, and $\eta_*$. We then have the following  possibilities for $\fp_*$:
\begin{itemize}
	\item If $n$ is odd then $\dim\fp_*=(n^2-1)/2$ and $\fp_*\cong \mathfrak{sl}_{(n+1)/2}\C\oplus\mathfrak{sl}_{(n-1)/2}\C\oplus\C$.
	\item If $n$, $\xi_*$, and $\eta_*$ are all even then $\dim\fp_*=n^2/2+1$ and $\fh_*\cong \mathfrak{sl}_{n/2+1}\C\oplus\mathfrak{sl}_{n/2-1}\C\oplus\C$.
	\item If $n$ is even and either $\xi_*$ or $\eta_*$ is odd then $\dim\fh_*=n^2/2-1$ and \smash{$\fp_*\cong \mathfrak{sl}_{n/2}\C\oplus\mathfrak{sl}_{n/2}\C\oplus\C$}.
\end{itemize}
We can see that for $n$ odd the subalgebras on the singular lines are all isomorphic, but for $n$ even the subalgebra on one of the four lines is not isomorphic to the other three. Note that the $\sigma_*$ are all inner automorphisms, and in particular their positive eigenspaces always contain an abelian summand. By inserting Wilson lines in 1-dimensional representations of these abelian summands along the associated singular lines we will be able to generate continuous families of $K$-matrices. Let's do this explicitly now.

\subsection{Asymptotic behaviour of elliptic $K$-matrices} 
\label{subsec:ellKmx}

Given the bulk propagator in holomorphic gauge, or equivalently the classical elliptic $r$-matrix, we can directly apply equation \eqref{eq:lmx} to determine the next-to-leading order contribution to the $K$-matrix in $\hbar$.

The term arising from self-interactions of the bulk Wilson line on the orbifold is determined by the representation of the bulk Wilson line, and is given by
\[ 
	\ell_0(z)=\frac{1}{2n}\sideset{}{'}
	\sum_{i,j\in\Z_n}\varepsilon^{-\zeta^{-1}ij}\,w_{i,j}(2z)\,t_{i,j}\,\sigma(t_{i,j})\,.
\]
Here, and in the sequel, the primed sum indicates that we remove the term proportional to the identity. If we choose the bulk Wilson line to be the fundamental representation\footnote{To avoid an anomaly on the bulk line, we must choose it to be in a representation of $\fg$ which lifts to a representation of the associated elliptic quantum group.} of $\mathfrak{sl}_n\C$ then we find that $t_{i,j}^2=\varepsilon^{-\zeta^{-1}ij}\,t_{2i,2j}$, so the self-interaction contribution becomes
\be 
\label{eq:elllmx0} 
	 \ell_0(z)=\frac{1}{2n}\sideset{}{'}
	 \sum_{i,j\in\Z_n}\varepsilon^{-2\zeta^{-1}ij+\xi i +\eta j}\,w_{i,j}(2z)\,t_{2i,2j}\,,
\ee
where we have used the action of $\sigma$ on $t_{i,j}$ given above.

\medskip

To determine the contribution of bulk-boundary interactions we first need to identify the generators of the abelian summand in each of the $\fh_*$. We will consider the cases where $n$ is odd/even separately.

The simplest case to deal with is $n$ odd, as in this case the four algebras $\fp_*$ are isomorphic. In terms of $n$, $\xi_*$, and $\eta_*$ the generator of $\C$ in $\fp_*$ is
\[ 
	Q_* = \frac{1}{n}\sideset{}{'}\sum_{i,j\in\Z_n}\varepsilon^{-2^{-1}(\zeta^{-1}ij+\xi_*i+\eta_*j)}\,t_{i,j}\,,
\]
where by $2^{-1}$ we mean the inverse of $2$ modulo $n$, which exists since $n$ is odd. Our representation of $\fp_*$ is given explicitly by \smash{$Q_*\mapsto {\tilde q}_*\in\C$}. The contribution of the bulk-boundary interactions is then
\[ 
	\delta\ell(z)=\frac{2}{n^2-1}\sum_{z^*}{\tilde q}_*\ \sideset{}{'}
	\sum_{i,j\in\Z_n}\,\varepsilon^{-2^{-1}(\zeta^{-1}ij+\xi_* i+ \eta_* j)}\,
	w_{i,j}(z-z^*)\,t_{i,j}\,.
\]
To combine this with the self-interaction contribution, it is helpful to first make the change of variables 
\[
	i \mapsto 2^{-1}i\qquad\text{and}\qquad j\mapsto 2^{-1}j
\]
in the sum in equation \eqref{eq:elllmx0}. Subsequently applying the identity
\[ 
	w_{2^{-1}i,2^{-1}j}(2z)=\frac{1}{2}\sum_{a_*,b_*\in\{0,1\}}\varepsilon^{ia_*+jb_*}w_{i,j}(z-z_*) 
\]
allows us to rewrite the self-interaction term as
\[
	\ell_0(z)=\frac{1}{4n}\sum_{a_*,b_*\in\{0,1\}}\sideset{}{'}\sum_{i,j\in\Z_n}		
	\varepsilon^{-2^{-1}(\zeta^{-1}ij + \xi_*i + \eta_*j)}\,w_{i,j}(z-z_*)\,t_{i,j}\,.
\]
Thus the full $\cO(\hbar)$ contribution to $K(z)\tau^{-1}$ is
\begin{equation}
	\ell(z) =\frac{1}{n}\sum_{z^*}q_*\sideset{}{'}\sum_{i,j\in\Z_n}\varepsilon^{-2^{-1}(\zeta^{-1}ij+\xi_*i+\eta_*j)}\,w_{i,j}(z-z_*)\,t_{i,j}\,,
\end{equation}
where
\[ 
	q_* = \left(\frac{1}{4}+\frac{2n}{n^2-1}{\tilde q}_*\right)\,.
\]
We delay a discussion of this solution until after we've dealt with the case where $n$ is even.

\medskip

The case $n=2m$ ($n$ even) is less straightforward as the four Lie subalgebras $\fp_*$ are no longer all isomorphic. The abelian generators in $\fp_*$ are easily expressed using the  formula
\[ 
	Q_* = \frac{1}{m}\sideset{}{'}\sum_{k,l\in\Z_n}t_{2k-\zeta\eta_*,2l-\zeta\xi_*}\,\varepsilon^{-2\zeta^{-1}kl}\,,
\]
where  again the primed sum indicates that we remove the term proportional to the identity. This appears only if \smash{$\xi_*\equiv\eta_*\equiv0~(\text{mod } 2)$}. Choosing 1 dimensional representations $Q_*\mapsto {\tilde q}_*$ for the four boundary Wilson lines, bulk-boundary interactions generate a contribution 
	%\begin{center}
	%	\begin{TAB}[5pt]{|c|c|}{|c|c|c|c|} 
	%		$\xi_*\equiv\eta_*\equiv 0 \ (2) $ & $Q_* = \frac{1}{m}\sum_{(k,l)\in\mathcal{I}_m}t_{2k,2l}\varepsilon^{-(2\zeta^{-1}kl+\xi_*+\eta_*l)}$ \\ 
	%		$\xi_*\equiv0\ (2)~\eta_*\equiv 1 \ (2) $ & $Q_*$ \\ 
	%		$\xi_*\equiv1\ (2)~\eta_*\equiv 0 \ (2) $ & $Q_*$ \\
	%		$\xi_*\equiv\eta_*\equiv 1 \ (2) $ & $Q_*$ \\ 
	%	\end{TAB}
	%\end{center}
	%It is possible to write down a general formula for all four of the above, however such an expression is not convenient for our subsequent calculations.
\[  
	\delta\ell(z)=\frac{1}{2m}\sum_{z^*}\sideset{}{'}\sum_{k,l\in\Z_m}p_*\,\varepsilon^{-2\zeta^{-1}kl}\,t_{2k-\zeta\eta_*,2l-\zeta\xi_*}\,w_{2k-\zeta\eta_*,2l-\zeta\xi_*}(z-z^*)\,.
\]
	%Here the \smash{${\tilde q}_*$} are proportional to the \smash{$q_*$}, with constants of proportionality depending on \smash{$\zeta$}, \smash{$m$}, \smash{$\xi_*$}, and \smash{$\eta_*$}.
where the \smash{$p_*$} are given by
\[ 
	{p}_* = \frac{2m}{m^2-(1-a_*)(1-b_*)}\varepsilon^{\zeta^{-1}(\zeta\xi_*-a_*)((\zeta\eta_*-b_*)/2)-\zeta^{-1}a_*b_*}{\tilde q}_*\,.
\]
Combining this with the self-interaction term shows that the full next-to-leading order contribution to $K(z) \tau^{-1}$ is
\begin{equation}
	\ell(z)=\frac{1}{m}\sum_{z^*}q_*\sideset{}{'}\sum_{k,l\in\Z_m}
	%\delta_{\xi_*\equiv\eta_*\equiv0~(2)}
	\varepsilon^{-2\zeta^{-1}kl}\,t_{2k-\zeta\eta_*,2l-\zeta\xi_*}\,w_{2k-\zeta\eta_*,2l-\zeta\xi_*}(z-z_*)
\end{equation}
in the even case, where now
\[
	q_* = \frac{1}{2}(1-a_*)(1-b_*)\varepsilon^{\zeta\xi_*(\eta_*/2)} 
	+ \frac{1}{2}p_*\,.
\]
plays the role of the `charges' of the boundary lines.	

Now that we've found the semi-classical contributions $\ell(z)$ for both even and odd $n$, let's discuss them. The formulas we have obtained agree with the semi-classical limits of the four-parameter elliptic $K$-matrices appearing in~\cite{Komori:1997yme}, as we demonstrate in appendix \ref{app:2}. The four continuous parameters of these solutions can be interpreted as the effective charges of each of the four boundary Wilson lines. (We will make a similar observation in the case of trigonometric $K$-matrices.)

From the point of view of gauge theory, a surprising feature of the $K$-matrices appearing in~\cite{Komori:1997yme} is that they are independent of the parameter playing the role of $\hbar$ in the bulk $R$-matrix. From the gauge theory perspective, this arises as follows. The next-to-leading order bulk Wilson line self-interaction term is inevitably proportional to $\hbar$ and contains no free parameters. However, we can cancel this contribution with an appropriate choice of the boundary charges $\tilde{q}_*$. Having achieved this cancellation, the remaining $\ell(z)$ is proportional to the boundary charges, which may now be rescaled to absorb $\hbar$. This ensures that the next-to-leading order contribution to the $K$-matrix can always be rescaled so as to remove $\hbar$. We anticipate that in the full perturbative expansion of the $K$-matrix, $\hbar$ may similarly be absorbed by the same shifts and rescalings of the boundary charges\footnote{The normalization of the $K$-matrix is fixed by the condition that the `Sklyanin determinant' be 1. This constraint depends on the $R$-matrix, and so generically we would expect the normalization to be $\hbar$ dependent.}. In this sense the $K$-matrix would be `independent' of $\hbar$.

It would be disappointing if the described independence was a general property of elliptic $K$-matrices that gauge theory obscures. Fortunately however, 
it appears to be special to the particular choice of (fundamental) representation we made for the bulk Wilson line. For generic representations we find that self-interactions of the bulk Wilson lines cannot be compensated by any choice of boundary parameters. For example, we claim that if we choose the bulk Wilson line to be in the adjoint representation\footnote{For this example to be valid, we need the adjoint representation of $\mathfrak{sl}_n\C$  to lift to a representation of the elliptic quantum group. We believe that~\cite{Cherednik:1985vs} demonstrates that the adjoint of $\mathfrak{sl}_n\C$ does indeed lift.}, then the $K$-matrix does depend on $\hbar$.

\medskip

The orbifold singularities $z_*$ should have significance beyond simply providing four free parameters for the $K$-matrix. To see this, recall that the category of representations of the (elliptic) quantum group has the property that $V_1(z_1)\otimes V_2(z_2)\ncong V_2(z_2)\otimes V_1(z_1)$, so the tensor product of representations  becomes non-commutative.  Nonetheless for generic values of the spectral parameters, the bulk $R$-matrix provides an non-trivial intertwiner
\[
	V_1(z_1)\otimes V_2(z_2) \ \buildrel{R}\over{\cong}\ V_2(z_2)\otimes V_1(z_1) \qquad \text{($z_1$, $z_2$ generic)}
\]
that fails when $z_1=z_2$ where the $R$-matrix becomes singular. The $R$-matrix thus provides us with a braided tensor category and, according to the general results of~\cite{Faddeev:1987ih}, any such category is equivalent to a category of representations of a Hopf algebra. Thus knowing the collection of $R$-matrices is really equivalent to knowing the elliptic quantum group itself.

For integrable systems in the presence of a boundary, the charges of the bulk quantum group will be broken to a (left) coideal subalgebra. In the rational case, this is the twisted Yangian $\cB(\fp,\fg)\subset\cY(\fg)$, but the appropriate algebraic structure in the elliptic case appears not to have been identified. The orbifold perspective strongly suggests that representations $W$ of this coideal subalgebra should have the property that the $K$-matrix provides an isomorphism
\[
	V(z) \otimes W \ \buildrel{K}\over{\cong}\ 
	V^\sigma(-z)\otimes W\qquad\text{($z$ generic)}
\]
between the tensor product of a representation $V(z)$ of the elliptic quantum group 
with $W$, and the tensor product where $V(z)$ is replaced by the representation $V^\sigma(-z)$ obtained by action of the orbifold $\Z_2$ on $V(z)$. Again, this isomorphism will fail when $z$ coincides with any of the four singular lines $z_*$.

\section{Trigonometric}
\label{sec:trig}

In this section we will discuss how to generate trigonometric $K$-matrices using gauge theory. Thus, in this section, we choose $C$ to be the cylinder $C=\R\times S^1=\C/\{z\sim z+2\pi i\}$. Again, we start by reviewing the method for generating trigonometric $R$-matrices described in~\cite{Costello:2017dso}. 

\subsection{Trigonometric solutions of the Yang-Baxter equation}
\label{sec:trigR}

To generate trigonometric solutions of the YBE from gauge theory, one must first find an isolated vacuum around which to perturb. This vacuum depends on a choice of boundary conditions as ${\rm Re}\,z\to\pm\infty$ in $C$. To retain the full topological invariance of the theory in $\Sigma$, these $C$ boundary conditions are independent of $\Sigma$. The na{\"i}ve boundary condition $A\to 0$ as ${\rm Re}\,z\to\pm\infty$ turns out to be too stringent, and instead Costello {\it et al.}~\cite{Costello:2017dso} require that the gauge field takes values in certain subspaces $\fg_{\pm}\subset\fg$ as $x\to\pm\infty$ respectively. For the boundary conditions to be consistent with gauge transformations, $\fg_\pm$ must in fact be Lie subalgebras of $\fg$ and the infinitesimal gauge transformations are also restricted to lie in these subalgebras as $z\to\pm\infty$. Furthermore, $\fg_\pm$ should be isotropic with respect to the $\fg$-invariant bilinear used to define the action, so as to eliminate a possible boundary term when varying $S[A]$. So as to impose the least restrictive condition possible, \cite{Costello:2017dso} choose $\fg_\pm$ each to be middle dimensional. Finally, to ensure the classical equations of motion to not admit deformations so that we are perturbing around an isolated minimum, one requires that \smash{$\fg_+\cap \fg_-=\varnothing$}.

Together these conditions require that $(\fg,\fg_-,\fg_+)$ form a Manin triple. If the invariant bilinear on $\fg$ is non-degenerate it provides an identification $\fg_-\cong\fg_+^*$.
%(In fact this identification endows $\fg_+$ with the structure of a Lie bialgebra.)
We can use this identification to extend a basis \smash{$\{t_a\}_{a=1}^{\dim\fg_+}$}  of $\fg_+$ to a basis of all of $\fg$ by writing \smash{$\{{\bar t}^a\}_{a=1}^{\dim\fg_+}$} for the dual basis viewed as a basis of $\fg_-$. In terms of this basis it's straightforward to write down the inverse of the bilinear
\[ c = \sum_a \big(t_a\otimes {\bar t}^a + {\bar t}^a\otimes t_a\big) = \sum_a t_a\otimes {\bar t}^a + \sum_a {\bar t}^a\otimes t_a = c_{+,-} + c_{-,+} \,,\]
where \smash{$c_{\pm,\mp}\in\fg_{\pm}\otimes\fg_{\mp}$}. Given a Manin triple, one finds 
\[ 
	r(z) = \frac{1}{e^z-1}c_{-,+} + \frac{e^z}{e^z-1}c_{+,-}
\]
as the classical $r$-matrix~\cite{Belavin:1982cybe}.

A generic simple Lie algebra cannot be given the structure of a Manin triple, since if $\fg=\fg_+\oplus\fg_-$ as a vector space and $\dim\fg_+=\dim\fg_-$, then $\fg$ itself must certainly be even dimensional. However, it is always possible to construct a Manin triple from any given simple Lie algebra $\fg_0$ by forming the direct sum of Lie algebras
\[ 
	\fg=\fg_0\oplus \widetilde\fh
\]
where $\widetilde\fh$ is a second copy of the Cartan subalgebra of $\fg_0$. Note that the Lie algebra $\fg$ is no longer simple since it has a non-trivial centre \smash{$\widetilde\fh$}. The $\fg$-invariant bilinear  used in defining the action of CWY theory is 
\[ 
	\langle~,~\rangle 
	= \langle~,~\rangle_{\fg_0} + \langle~,~\rangle_{\widetilde \fh}\,,
	\]
where \smash{$\langle~,~\rangle_{\fg_0}$} denotes a symmetric invariant bilinear on $\fg_0$ proportional to the Killing form, and \smash{$\langle~,~\rangle_{\widetilde \fh}$} denotes its restriction to the Cartan. To determine one of the  Lagrangian subalgebras, say $\fg_+$ one picks a decomposition $\fg_0 = \mathfrak{n}_-\oplus\fh\oplus \mathfrak{n_+}$, where $\fh$ is a choice of Cartan and $\mathfrak{n}_\pm$ are the subalgebras of positive and negative root spaces for a given base. Then
\[
	\fg_+ = \mathfrak{n}_+\oplus \left\{(H,i\widetilde H)\,| \, H\in\fh\right\}\,.
\]
This definition makes a choice of identification between the Cartan $\fh$ of $\fg_0$ and the centre $\widetilde\fh$ of $\fg$. The other Lagrangian subalgebra $\fg_-$ is similarly defined as
\[
	\fg_- =\mathfrak{n}_-\oplus \big\{(H,iM(H))\, | \,H\in\fh\big\}\,,
\]
where $M$  allows for a different identification between $\fh$ and $\widetilde\fh$ in $\fg_-$.  The linear map $M$ must be orthogonal to ensure that  $\fg_-$ is isotropic, and $+1$ must not be an eigenvalue of $M$ if $\fg_+$ and $\fg_-$ are to be disjoint.

To write the $r$-matrix in the basis adapted to this Manin triple, we let $\{e_\mu,f_\mu\}_{\mu\in\Phi_+}\cup\{h_\mu\}_{\mu\in\Delta}$ be the standard Chevalley basis of $\fg_0$ with respect to our choice of base $\Delta$. Here $\Phi_+$ denotes the set of positive roots for this base. By adjoining $\{{\widetilde h}_\mu\}_{\mu\in\Delta}$ we can extend this to a basis of $\fg$. We write $\kappa$ for the restriction of our $\fg_0$-invariant bilinear to $\fh$, and use it to raise and lower indices in both copies of the Cartan. Finally we define the antisymmetric bivector $A$ by
\[ 
A\kappa = (M+1)(M-1)^{-1}
\]
where $M\in O(\widetilde\fh)$ is the orthogonal transformation introduced above. Then the classical $r$-matrix becomes
\[ 
\begin{aligned}
	r(z) = &\frac{1}{e^z-1}\sum_{\mu\in\Phi^+} f_\mu\otimes e_{\mu} 
	+ \frac{e^z}{e^z-1}\sum_{\mu\in\Phi_+}e_\mu\otimes f_\mu  \\ 
	&\ +\frac{1}{2}\frac{e^z+1}{e^z-1}\sum_{\mu,\nu\in\Delta} \kappa^{\mu\nu}(h_\mu\otimes h_\nu + h_\nu\otimes h_\mu) + \frac{1}{2}\sum_{\mu,\nu\in\Delta}A^{\mu\nu}(h_\mu\otimes h_\nu - h_\nu\otimes h_\mu)  \\ 
	&\ -\frac{i}{2}\sum_{\mu,\nu\in\Delta}(\kappa^{\mu\nu} - A^{\mu\nu})h_\mu\otimes {\widetilde h}_\nu + \frac{i}{2}\sum_{\mu,\nu\in\Delta}(\kappa^{\mu\nu} + A^{\mu\nu}) {\widetilde h}_\mu\otimes h_\nu
\end{aligned} 
\]
in this basis.

\subsection{Trigonometric solutions of the boundary Yang-Baxter equation}
\label{sec:TrigBoundary}

We now consider the theory on the orbifold $\widetilde M = (\Sigma\times\R\times S^1)/\Z_2$. As usual we take the $\Z_2$ to act as
\[ 
	Z:(x,y,z,\bar z)\mapsto (-x,y,-z,-\bar z)\,.
\]
This map swaps ${\rm Re}\,z = +\infty$ and ${\rm Re}\,z = -\infty$. Above, we saw that in the trigonometric case, different boundary conditions are imposed on $A$ in these two limits, with $A$ lying in $\fg_\pm$ as ${\rm Re}\,z\to\pm\infty$, respectively. To handle this situation we must generalise our orbifold construction so that the lift of the $\Z_2$ action to the adjoint bundle also exchanges the different boundary conditions.

The simplest way to achieve this is to require that the involutive automorphism $\sigma:\fg\to\fg$ exchanges the subalgebras ${\fg}_-$ and ${\fg}_+$, {\it i.e.} we require $\sigma(\fg_+) = \fg_-$. In terms of the basis \smash{$\{t_a\}_{a=1}^{\dim\fg_+}$} of $\fg_+$ and the associated dual basis \smash{$\{{\bar t}^a\}_{a=1}^{\dim\fg_+}$} of $\fg_-$, the +1 eigenspace of $\sigma$ is spanned by \smash{$\{u_a\}_{a=1}^{\dim \fg_+}$} where
\[ 
	u_a = t_a + \sigma(t_a)\,.
\]
It is then straightforward to write down the semi-classical contribution to the $K$-matrix using equation \eqref{eq:lmx}. It is given by
\[ 
\begin{aligned}
\ell(z) = &\frac{1}{2}\frac{1}{e^{2z}-1}c_-^\sigma\otimes{\bf 1}\otimes{\bf 1} + \frac{1}{2}\frac{e^{2z}}{e^{2z}-1}c_+^\sigma\otimes{\bf 1}\otimes{\bf 1}  \\ 
&\qquad +\ \frac{2}{e^z-1}\sum_a\tbar^{\,a}\otimes u_a\otimes{\bf 1} + \frac{2e^z}{e^z-1}\sum_a\sigma(\tbar^{\,a})\otimes u_a\otimes{\bf 1} \\ 
&\qquad\qquad- \ \frac{2}{e^z+1}\sum_a\tbar^{\,a}\otimes{\bf 1}\otimes u_a + \frac{2e^z}{e^z+1}\sum_a\sigma(\tbar^{\,a})\otimes{\bf 1}\otimes u_a\,,
\end{aligned} 
\]
where the three factors in the tensor product correspond to the bulk and two boundary Wilson lines, and where
\[ \begin{aligned}
c_-^\sigma = {\rm gl}\big(\sigma_1(c_{+,-})\big) = \sum_a \sigma(t_a) \tbar^{\,a}
= \sum_a \tbar^{\,a} \sigma(t_a) \,, \\
c_+^\sigma = {\rm gl}\big(\sigma_1(c_{-,+})\big) = \sum_a \sigma(\tbar^{\,a}) t_a = \sum_a t_a \sigma(\tbar^{\,a})\,. \\
\end{aligned} 
\]
and glues the two  entries of $c_{+,-}$ or $c_{-,+}$ together, after acting with $\sigma$ on the first entry. Note that for $\sigma$ to be a symmetry of the theory it is essential that $\sigma$ preserves our chosen invariant bilinear on $\fg$, which we have used in the above. Since $\fg$ is not semisimple this is not an immediate consequence of the fact that $\sigma$ is an automorphism.

The classification of involutive automorphisms swapping $\fg_+$ and $\fg_-$ for a Manin triple $(\fg,\fg_+,\fg_-)$ follows from standard Lie theory and is given in appendix \ref{app:3}. The outcome is that such $\sigma$ exist only for certain choices of $M$, and are completely determined by a choice of involutive automorphism of the simple Lie algebra $\fg_0$. We denote this involution by $\chi$. The data we use in defining $\chi$ is an involutive automorphism $\gamma$ of the Dynkin diagram of $\fg_0$, together with an element $\lambda\in\fh$ which is invariant under then natural action of $\gamma$ on the Cartan. Explicitly
\[ 
	\chi = \exp({\rm ad}_\lambda)\circ\Gamma\circ\omega 
\]
where
\[ 
\omega: (e_\mu,f_\mu,h_\mu) \mapsto (f_\mu,e_\mu,-h_\mu)
\qquad\text{for $\mu\in\Delta$} 
\]
is the Chevalley involution, and where $\Gamma$ extends the action of $\gamma$ to all of $\fg_0$ by
\[ 
	\Gamma: (e_\mu,f_\mu,h_\mu) \mapsto 	
	(e_{\gamma(\mu)},f_{\gamma(\mu)},h_{\gamma(\mu)})
	\qquad\text{for  $\mu\in\Delta$} \,.
\]
We get an associated solution of the  boundary Yang-Baxter equation whenever
\be
\label{Mchiinvol} 
	(M\circ\chi|_{\widetilde\fh})^2={\bf 1}_{\widetilde\fh}\,.
\ee
where $\chi|_{\widetilde\fh}$ is the restriction of $\chi$ to the Cartan. The action of $\sigma$ on $\widetilde\fh$ is given by $M\circ\chi|_{\widetilde\fh}$. In particular, if $\gamma$ is the identity, \eqref{Mchiinvol} implies that $M$ itself is involutive. In this case, since $M$ has no +1 eigenvalue, $M$ is fixed to be $M = -{\bf 1}_{\widetilde h}$.

\medskip

We now construct explicit examples of such automorphisms and their associated $K$-matrices for some low dimensional $\fg$.

\subsection{Examples of trigonometric solutions}
\label{sec:trigexamples1}

We begin by considering $\fg=\mathfrak{sl}_2\C\oplus\C$ with basis $\{e,f,h,{\widetilde h}\}$. The Dynkin diagram of $\mathfrak{sl}_2\C$ admits only the trivial diagram automorphism $\gamma={\rm id}$, so
\[ 
	\chi = \exp({\rm ad}_{\lambda h})\circ\omega 
\]
for some $\lambda\in\C$. This involutive automorphism can be realised explicitly as conjugation by
\begin{equation}
\label{tausl2}
\tau = i\begin{pmatrix}
0 & \Lambda^{-1/2} \\ \Lambda^{1/2} & 0
\end{pmatrix} \in SL_2\C
\end{equation}
where $\Lambda = e^{2\lambda}$. As $\gamma$ is trivial we must have $M=-1$:  indeed, this is the only orthogonal transformation of a 1 dimensional space with no +1 eigenvalue. Thus $\sigma$ acts on $\widetilde h$ trivially. The positive eigenspace $\fp$ of $\sigma$ is then an abelian Lie algebra of dimension 2 spanned by
\begin{equation}
\label{boundary+sl2}
 	\left\{ \Lambda^{-1/2}e + \Lambda^{1/2} f = 
	\begin{pmatrix}
		0 & \Lambda^{-1/2} \\ 
		\Lambda^{1/2} & 0
	\end{pmatrix},\  {\widetilde h}\right\}\,.
\end{equation}
We may insert boundary Wilson lines at $u=e^z=\pm 1$ in representations of this algebra.

\medskip 

To define a $K$-matrix we need to make appropriate choices of representations for the bulk and boundary Wilson lines. We will take the bulk Wilson line to be in the tensor product of the fundamental of $\mathfrak{sl}_2\C$ with the charge $s$ representation of $\widetilde\fh = \C$. Bulk $R$-matrices associated to Wilson lines with non-vanishing charges for $\widetilde\fh$ correspond to a generalization of the 6-vertex model with non-vanishing horizontal and vertical fields~\cite{Costello:2017dso}. Such representations were referred to as `inadmissible' in~\cite{Costello:2018gyb}.

We will choose the boundary Wilson lines at $u=\pm1$ to be in 1 dimensional representations of the abelian Lie algebra~\eqref{boundary+sl2} labelled by charges $\{q_\pm,r_\pm\}$, respectively. The next-to-leading contribution to the $K$-matrix is then 
\[ 
\begin{aligned}
	k(z)\tau^{-1} = -\frac{1}{8}\frac{e^{2z}+1}{e^{2z}-1}(1-s^2)
	 + \frac{1}{2}\frac{e^z+1}{e^z-1}r_1s + \frac{1}{2}\frac{e^z-1}{e^z+1}r_2s 
	  -\frac{i}{4}(2r_+ + 2r_- + s)h\\
	+\ \Lambda^{-1/2}\left(q_+\frac{e^z}{e^z-1} + q_-\frac{e^z}{e^z+1}\right) e
	+ \Lambda^{1/2}\left(q_+\frac{1}{e^z-1} - q_-\frac{1}{e^z+1}\right)f\,.
\end{aligned} 
\]
The first three terms here are proportional to the identity, and so can be absorbed into the overall normalization of the $K$-matrix. The terms that cannot be removed in this way are 
\begin{multline*} 
	-\frac{i}{4}(2r_+ + 2r_- + s)h +\frac{1}{2}e^{z/2}\Lambda^{-1/2}
	\left(q_+\frac{1}{\sinh(z/2)} + q_-\frac{1}{\cosh(z/2)}\right)e  \\
	+\  \frac{1}{2}e^{-z/2}\Lambda^{1/2}
	\left(q_+\frac{1}{\sinh(z/2)} + q_-\frac{1}{\cosh(z/2)}\right)f\,. 
\end{multline*} 
The full $K$-matrix can then be expressed as 
\[ 
\begin{aligned}
	K(z)\tau^{-1} =& F(z;\hbar)\left(\sinh(z) - \frac{i\hbar}{4}\sinh(z) (s+2r)h
	+ \hbar e^{z/2}\Lambda^{-1/2} q\sinh(\xi/2+z/2) e \right.\\
	&\left.\phantom{\frac{1}{1}}\qquad\qquad + \hbar e^{-z/2}\Lambda^{1/2}  q\sinh(\xi/2-z/2)f\right) 
	+ \mathcal{O}(\hbar^2)
\end{aligned} 
\]
where the dressing function
\[ 
	F(z;\hbar) = \frac{1}{\sinh(z)}\left(1 - \frac{\hbar}{8}(1-s^2)\coth(z)
	+ \frac{\hbar}{2}r_+s\coth(z/2) + \frac{\hbar}{2}r_-s\tanh(z/2)\right)\,,
\]
and where $(q,\xi,r)\in\C$ are defined by $q_+ = q\sinh(\xi/2)$, $q_- = q\cosh(\xi/2)$ and $r = r_+ + r_-$. Equivalenty, in the standard representation of $\mathfrak{sl}_2$ and multiplying through on the right by $\tau$ as in~\eqref{tausl2}, we have
\[ 
	K(z) = iF(z;\hbar)\begin{pmatrix} 
	\hbar q\, e^{z/2}\sinh(\xi/2+z/2) & 
	\Lambda^{-1/2}(1-i\hbar(s+2r)/4)\sinh(z) \\ 
	\Lambda^{1/2}(1+i\hbar(s+2r)/4)\sinh(z) & 
	\hbar q\,e^{-z/2}\sinh(\xi/2-z/2)
	\end{pmatrix} \,
	+ \, \mathcal{O}(\hbar^2)\,.
\]
To compare this with $K$-matrices appearing in the literature, it is convenient to first symmetrize the $R$-matrix, conjugating it by
\[ 
	i\begin{pmatrix} 0 & e^{z/4} \\ e^{-z/4} & 0 
	\end{pmatrix}
\]
as discussed in~\cite{Costello:2017dso}. After performing the corresponding transformation of the $K$-matrix, we arrive at 
\[ 
	K(z) = iF(z;\hbar)\begin{pmatrix}
	\hbar q\, \sinh(\xi/2 - z/2) & \Lambda^{1/2}(1+i\hbar (s+2r)/4)\sinh(z) \\
	\Lambda^{-1/2}(1-i\hbar (s+2r)/4)\sinh(z) & \hbar q\, \sinh(\xi/2 + z/2)
	\end{pmatrix}\,.
\]
This matches the expression appearing in~\cite{deVega:1993xi}.

\bigskip

As a second example, let's also consider $\fg = {\mathfrak{sl}_3\C}\oplus\C^2$. For the sake of simplicity we restrict our attention to involutive automorphisms for which $\chi$ is inner. Since the Chevalley involution of $\mathfrak{sl}_3\C$ is in fact an outer automorphism, we must choose $\gamma$ to be the non-trivial outer automorphism of $\mathfrak{sl}_3\C$. This simply swaps the two simple roots, {\it i.e.} $\gamma = (1\,2)$. Then
\[ 
	\chi = {\rm ad}_\lambda\circ\Gamma\circ\omega
\]
as above. In this case, since $\lambda$ is $\gamma$-invariant it has 1 degree of freedom, so $\lambda = a(h_1 + h_2)$ for some $a\in\C$. Acting on the Chevalley generators $\{e_i,f_i,h_i\}_{i\in{0,1}}$ of $\mathfrak{sl}_3\C$,  $\chi$ is given by
\[ 
	h_1\mapsto -h_2\,,\qquad 
	e_1\mapsto A^{-1}f_2\,,\qquad 
	e_2\mapsto A^{-1}f_1\,,
\]
where $A = e^a$. The action of $\chi$ on the remaining generators is determined by the fact it is involutive. Explicitly, $\chi$ can be realised as conjugation by the matrix
\[ 
\tau = \begin{pmatrix}
				0 & 0 & A \\
				0 & 1 & 0 \\
				A^{-1} & 0 & 0
			\end{pmatrix} \in PSL_3\C\,.
\]
Since $\gamma$ is no longer trivial there exist non-trivial $M\in O(\widetilde\fh)$ for which
\[ 
	(M\circ\chi|_{\widetilde\fh})^2={\mathbf 1}_{\widetilde\fh}\,.
\]
In this example, this constraint fixes $M$ to be a non-trivial (complex) rotation. This includes the possibility that $M=-{\bf 1}_{\widetilde\fh}$ and, for the sake of simplicity, we will start by concentrating on this case, which is the standard case. Under this assumption, $\sigma$ acts trivially on $\widetilde\fh$, and its positive eigenspace is the direct sum $\fp=\fp_0\oplus\widetilde\fh$, where
\[
	\fp_0 = \text{span}_\C\left\{A^{1/2}e_1 + A^{-1/2}f_2\,,\ 
	A^{1/2}e_2 + A^{-1/2}f_1\,,\ 
	A e_\gamma + A^{-1}f_{\gamma}\,,\ h_2-h_1\right\}\,.
\]
Here $\gamma = \alpha_{(1)}+\alpha_{(2)}$ for $\alpha_{(1)}$ and $\alpha_{(2)}$ the simple roots of $\mathfrak{sl}_3\C$, so $e_\gamma=[e_1,e_2]$ and $f_\gamma=[f_2,f_1]$. $\fp_0$ is a subalgebra of $\mathfrak{sl}_3\C$ isomorphic to $\mathfrak{sl}_2\C\oplus\C$. Note that 
\[ 
	h_2 - h_1 + 3Ae_\gamma + 3A^{-1}f_\gamma
\]
is the generator of the abelian summand of this subalgebra.

\medskip 

We are now in a position to compute the leading order contribution to the corresponding $K$-matrix. We will take the bulk Wilson line to be in the fundamental representation of $\mathfrak{sl}_3\C$ with no charge under the $\widetilde \fh$ factor, {\rm i.e.} we will take the bulk Wilson line to be admissible. We will take the boundary Wilson lines at $u=\pm1$ to be in 1 dimensional representations of $\fp_0$ of charge $q_\pm$ respectively, and to have no charge under $\widetilde \fh$. (These assumptions are made to simplify the resulting $K$-matrix.)

Under these assumptions we can write down the next-to-leading order contribution to the $K$-matrix explicitly. By repeating similar calculations as for $\fg_0=\mathfrak{sl}_2\C$ we arrive at the  expression
\[ 
	k(z)\tau^{-1} = \frac{1}{\sinh z} 
	\begin{pmatrix}
		a & 0 & b \\
		0 & -2a & 0 \\
		c & 0 & a
	\end{pmatrix}\,,
\]
where 
\[
\begin{aligned}
	a &= -\frac{1}{12}(\cosh z + 2q_+ \cosh^2(z/2) + 2q_-\sinh^2(z/2) )\\
	b &= \frac{A}{4}e^{z/2}(e^z/2 + 2q_+\cosh(z/2) + 2q_-\sinh(z/2)) \\
	c &= \frac{A^{-1}}{4}e^{-z/2}(e^{-z/2} + 2q_+\cosh(z/2) - 2q_-\sinh(z/2))
\end{aligned}
\]
This form is highly suggestive: It is clear that, with an appropriate choice of the parameters $q_+$ and $q_-$, we can eliminate the contribution from the self-interaction of the bulk Wilson line. In particular, if we introduce the parameters $q_1$ and $q_2$ by $2q_+ = q_1 + q_2 -1$ and $2q_- = q_1-q_2-1$, the above $k$-matrix simplifies to
\[ 
	k(z)\tau^{-1} = \frac{1}{\sinh z}\begin{pmatrix}
	-\frac{1}{12}(q_1\cosh z + q_2) & 0 & 
	\frac{A}{4}e^{z/2}(q_1e^{z/2} + q_2e^{-z/2}) \\
	0 & \frac{1}{6}(q_1\cosh z + q_2) & 0 \\
	\frac{A^{-1}}{4}e^{-z/2}(q_1e^{-z/2} + q_2e^{z/2}) & 0 & 
	-\frac{1}{12}(q_1\cosh z + q_2)
	\end{pmatrix}\,.
\]
This cancellation of the bulk self-interactions with boundary charges was observed in the elliptic case also, and seems to be a common feature of $K$-matrices associated with the $A_{n-1}\cong\mathfrak{sl}_n\C$ family of classical Lie algebras, at least in the fundamental representation.  We emphasise again that this is not a generic feature of all $K$-matrices, indeed there exist even rational $K$-matrices which depend explicitly on $\hbar$.
%\red{It would be good to check that in the rational case non-trivial representations of $\mathfrak{sl}_n\C$ do indeed give $\hbar$-dependent $K$-matrices.}

From the above formula we arrive the  final expression for the $K$-matrix
\[ 
\begin{aligned}
	&K(z) = \frac{i}{\sinh z}\ \times \\
	&\scalemath{0.85}{\begin{pmatrix}
	\frac{\hbar}{4}(q_1e^z+q_2) & 0 & 
	A\big(\sinh z-\frac{\hbar}{12}(q_1\cosh z + q_2)\big) \\
	0 & \sinh z + \frac{\hbar}{6}(q_1\cosh z + q_2) & 0 \\
	A^{-1}\big(\sinh z-\frac{\hbar}{12}(q_1\cosh z + q_2)\big) & 0 & 
	\frac{\hbar}{4}(q_1e^{-z}+q_2)\end{pmatrix}}+ \mathcal{O}(\hbar^2)\,.
\end{aligned}
\]
This coincides with one of the three families solutions constructed for $\fg=\mathfrak{sl}_3\C$ in~\cite{LimaSantos:2002uh}. 

%\red{I need to check this more carefully. I will do so after I've finished the remaining two examples as I have a nasty feeling that the $R$-matrix still needs to be conjugated into the correct form.} 

\medskip

It is conspicuous that we are missing the remaining two families of trigonometric $K$-matrices obtained in~\cite{LimaSantos:2002uh}. Studying these solutions carefully one finds that the analogue of the classical limit for these $K$-matrices is not $z$-independent. This motivates us to consider a generalization of the construction presented so far.

\subsection{Generalization to $z$-dependent automorphisms}
\label{sec:trigk2}

To accommodate $K$-matrices with $z$-dependent classical limits, we choose to lift the action of the map $Z:(x,y,z,\bar{z})\mapsto (-x,y,-z,-\bar{z})$ to the adjoint bundle of the gauge theory not by simultaneously acting on its fibres with a constant involutive automorphism, but rather using a bundle morphism $\sigma_Z: {\rm Ad}\,P\to{\rm Ad}\,P$. We choose this bundle morphism such that the following diagram commutes:
\[
\begin{tikzcd}
	{\rm Ad}\,P \arrow{d}{\pi} \arrow{r}{\sigma_Z} & {\rm Ad}\,P \arrow{d}{\pi} \\
	M \arrow{r}{Z}& M
\end{tikzcd}\,,
\]
where $\sigma_Z$ sends the fibre over $(x,y,z,\bar z)$ to fibre over $(-x,y,-z,-\bar{z})$ whilst also acting with an involutive automorphism $\sigma(z)$ of $\fg$. If the adjoint bundle had non-trivial topology then this automorphism $\sigma(z)$ would not exist globally, but in the trigonometric case the adjoint bundle is trivial. The bundle morphism is then completely determined by the holomorphic map $\sigma(z)$ from $\C^*$ to the automorphism group of $\fg$. To preserve the vacuum $A=0$, this automorphism must be independent of $\Sigma$, but in contrast to the elliptic case considered earlier, in the trigonometric case $\sigma$ may vary holomorphically with $z$. (We show in appendix~\ref{app:EllipticBundleMorphisms} that nothing new arises if we allow this generalisation in the elliptic case.) For $\sigma_Z$ to be involutive $\sigma(z)$ must obey $\sigma(z)\sigma(-z) = {\bf 1}$. Note that for $S[A]=S[\sigma_Z(Z^*\!A)]$, we require that $\sigma_Z$ preserves the $\fg$-invariant bilinear on the fibres.

Note that in the trigonometric case we still require our automorphism to exchange the boundary conditions at $\pm\infty$. We also require that our automorphism map non-singular field configurations to non-singular field configurations.

To compute the next-to-leading order contribution to the $K$-matrix for a $z$-dependent automorphism $\sigma(z)$ we proceed exactly as before. In holomorphic gauge, the propagator is 
\[
	P_{xy}(w;w') = \delta^{(2)}_\Sigma\!({\bf r} - {\bf r'})\,r(z-z') 
	+ \delta^{(2)}_\Sigma\!\big({\bf r} - Z({\bf r'})\big)\,\sigma_2(z')(r(z+z')) \,,
\]
the only difference being that $\sigma$ is now $z$-dependent. We find that the classical $K$-matrix is given by
\[ 
	K(z)\tau^{-1}(z) = \frac{1}{2}{\rm gl}\big(\sigma_2(z)(r(2z)\big)|_V 
	+ 2 \sum_{z_*}r(z-z_*)|_{V\otimes W_*}\,,
\]
with the two terms again coming from evaluating the Feynman diagrams describing self-interaction of the bulk Wilson line, and interactions between the bulk and boundary Wilson lines.

\subsection{Examples of $K$-matrices associated to $z$-dependent automorphisms}

In appendix~\ref{app:loopalg} we consider how $z$-dependent bundle morphisms can be constructed from automorphisms of the loop algebra $L\fg$. The simplest example of a $z$-dependent automorphism arises for $\fg = \mathfrak{sl}_2\C\oplus\C$, and acts on the $\mathfrak{sl}_2\C$ summand as conjugation by
\[ 
	\tau(z) = \begin{pmatrix} e^{z/2} & 0 \\ 0 & e^{-z/2} \end{pmatrix} \,.
\]
This automorphism corresponds to the non-trivial transposition of the Dynkin diagram of the affine algebra $\widehat{\mathfrak{sl}_2\C}$ (see appendix~\ref{app:loopalg}). It extends to act on $\widetilde \fh \cong \C$ as $\widetilde h\mapsto -\widetilde h$. The Lie subalgebras of $\fg$ that survive at the fixed points $u=\pm1$ are isomorphic to $\mathfrak{sl}_2\C$ and $\C = \langle h\rangle_\C$ respectively\footnote{By replacing $u$ by $-u$ in $\tau(z)$ we can swap the subalgebras at $\pm1$, which will generate a $K$-matrix essentially equivalent to the one constructed here.}. To generate solutions of the boundary Yang-Baxter equation without boundary degrees of freedom we must insert boundary Wilson lines in 1 dimensional representations of these algebras. There exist no such representations for $\mathfrak{sl}_2\C$, so we can only insert a Wilson line at $u=-1$ in the representation $h\mapsto q\in\C$. We will take the reflecting bulk Wilson line to be in the fundamental representation of $\mathfrak{sl}_2\C$ and to be uncharged under $\widetilde\fh$.

After some algebra, one arrives at the $K$-matrix\footnote{For ease of comparison to the $K$-matrices in~\cite{deVega:1993xi} we have included $\hbar$ in the argument of the hyperbolic functions here. Nonetheless we emphasize that this $K$-matrix is valid only to next-to-leading order in $\hbar$.}
\[ 
	K(z) = \frac{1}{\cosh{z/2}}
	\begin{pmatrix} 
		e^{z/2}\cosh(2\hbar q +z/2) & 0 \\ 
		0 & e^{-z/2}\cosh(2\hbar q - z/2)
	\end{pmatrix} + \mathcal{O}(\hbar^2)\,.
\]
To symmetrise the $R$-matrix we conjugate it by
\[
	 i\begin{pmatrix} 0 & e^{z/4} \\ e^{-z/4} & 0 \end{pmatrix}\,,
\]
under which $K(z)$ transforms to
\[ K(z) = \frac{i}{\cosh{z/2}}\begin{pmatrix} \sinh(-\frac{\pi i}{2} + 2\hbar q + z/2) & 0 \\ 0 & \sinh(-\frac{\pi i}{2} + 2\hbar q - z/2)\end{pmatrix} + \mathcal{O}(\hbar^2)\,.\]
In this form it is clear that this solution belongs to the same family of $K$-matrices in~\cite{deVega:1993xi} as those constructed using a constant $\sigma$. However this solution is expanded around a slightly different classical limit.

\medskip

Now we apply this construction to $\mathfrak{sl}_3\C\oplus\C^2$, which will allow us to generate less trivial examples. Consider the $z$-dependent automorphism acting on $\mathfrak{sl}_3\C$ as conjugation by
\[ \tau(z) = i\begin{pmatrix}
e^z & 0 & 0 \\
0 & 0 & A \\
0 & A^{-1} & 0
\end{pmatrix}\,.\]
This corresponds to the non-trivial transposition of the Dynkin diagram of the affine algebra $\widehat{\mathfrak{sl}_3\C}$ that swaps the 0\textsuperscript{th} and 1\textsuperscript{st} simple roots. (One can construct a similar automorphism for the transposition swapping the 0\textsuperscript{th} and 2\textsuperscript{nd} roots.) Under the simplifying assumption that $M=-1$, this automorphism extends to the whole of $\fg$ by acting on $\widetilde\fh$ as $\widetilde{h}_1\mapsto -\widetilde{h}_1-\widetilde{h}_2$ and $\widetilde{h}_2\mapsto\widetilde{h}_2$. The subalgebras that survive at the points $u=\pm1$ are both isomorphic to the direct sum of $\langle\widetilde{h}_2\rangle_\C\subset\widetilde\fh$ with $\mathfrak{sl}_2\C\oplus\C\subset\fg_0$. The abelian summand of $\mathfrak{sl}_2\C\oplus\C$ here is generated by
\[ 
	Q=2h_1 + h_2 \pm (3A e_2 + 3A^{-1}f_2)\,.
\]

To construct a $K$-matrix, we will take the reflecting bulk Wilson line to be in the fundamental representation and uncharged under $\widetilde \fh$. We  similarly take our boundary Wilson lines to be uncharged under $\widetilde\fh$, and as usual will assume them to be in a 1 dimensional representations of $\mathfrak{sl}_2\C\oplus\C\subset\fg_0$. Thus our 1 dimensional representations at $u=\pm1$ map $Q\to q_\pm$, respectively. A somewhat tedious calculation shows that the leading order contribution to the $K$-matrix then takes the form
\[ 
	k(z)\tau^{-1}(z) = \frac{1}{\sinh z}
			\begin{pmatrix}
				-2a & 0 & \ 0 \ \\
				0 & a & \ b\  \\
				0 & c & \ a\ 
			\end{pmatrix}\,,
\] 
where
\[
\begin{aligned}
	a &= -\frac{1}{12}(\cosh z + q_+ \cosh^2(z/2) + q_-\sinh^2(z/2) )\\
	b &= \frac{A}{4}e^{z/2}(e^{-z/2} + q_+\cosh(z/2) - q_-\sinh(z/2)) \\
	c &= \frac{A^{-1}}{4}e^{-z/2}(e^{z/2} + q_+\cosh(z/2) + q_-\sinh(z/2))\,.
\end{aligned}
\]
If we introduce the parameters $q_1$ and $q_2$ by $q_+ = q_1 + q_2 -1$ and $q_- = q_1-q_2-1$, the above $k$-matrix simplifies to
\[ 
k(z)\tau^{-1} = \frac{1}{\sinh z}\begin{pmatrix}
\frac{1}{6}(q_1\cosh z + q_2) & 0 & 
0 \\
0 & -\frac{1}{12}(q_1\cosh z + q_2) & \frac{A}{4}e^{z/2}(q_1e^{-z/2} + q_2e^{z/2}) \\
0 & \frac{A^{-1}}{4}e^{-z/2}(q_1e^{z/2} + q_2e^{-z/2}) & 
-\frac{1}{12}(q_1\cosh z + q_2)
\end{pmatrix}\,.
\]
Note the by now familiar fact that the self-interactions of the bulk $R$-matrix can be eliminated with a suitable choice of boundary parameters. From the above formula we arrive the  final expression for the $K$-matrix
\[ 
\begin{aligned}
&K(z) = \frac{i}{\sinh z}\ \times \\
&\scalemath{0.85}{\begin{pmatrix}
	e^z\big(\sinh{z}+\frac{\hbar}{6}(q_1\cosh z + q_2)\big) & 0 & 
	0 \\
	0 & \frac{\hbar}{4}(q_1 + q_2e^z) & A\big(\sinh{z}-\frac{\hbar}{12}(q_1\cosh z + q_2)\big) \\
	0 & A^{-1}\big(\sinh{z}-\frac{\hbar}{12}(q_1\cosh z + q_2)\big) & 
	\frac{\hbar}{4}(q_1 + q_2e^{-z})\,.
	\end{pmatrix}} + \mathcal{O}(\hbar^2)
\end{aligned}
\]
This coincides with one of the two remaining families of solutions constructed for $\fg=\mathfrak{sl}_3\C$ in~\cite{LimaSantos:2002uh}. We also claim that choosing
\[ 
\tau(z) = i\begin{pmatrix}
			0 & A & 0 \\
			A^{-1} & 0 & 0 \\
			0 & 0 & z^{-1}
	\end{pmatrix}
\]
allows us to generate the final family of solutions.

\appendix

\section{Asymptotic behaviour of elliptic $K$-matrices in the $\hbar\to0$ limit}
\label{app:2}
In this appendix we take the semi-classical limit of the elliptic $K$-matrix appearing in~\cite{Komori:1997yme}, and verify that the next-to-leading order contribution matches the result arrived at using gauge theory. In this appendix we will always take $\zeta=1$.

Expressed in terms of the basis $\{t_{i,j} = B^iA^{-j}\}_{i,j\in\Z_n}$ of ${\rm End}(\C^n)$, where $t_{0,0}={\bf 1}$, Belavin's symmetric $R$-matrix is\footnote{We have expressed the $R$-matrix in terms of the natural parameters appearing in the gauge theory description, and have picked a normalization in which its classical limit is the identity.}
\[ 
	R(z) = \sum_{i,j\in\Z_n}\varepsilon^{-ij} \frac{\theta\!\left[\begin{smallmatrix} 1/2 + i/n \\ 1/2 - j/n \end{smallmatrix}\right]\!(z+\hbar/n|\tau)\,\theta\!\left[\begin{smallmatrix}1/2 \\ 1/2\end{smallmatrix}\right]\!(\hbar/n|\tau)}{\theta\!\left[\begin{smallmatrix}1/2 \\ 1/2\end{smallmatrix}\right]\!(z|\tau)\,\theta\!\left[\begin{smallmatrix} 1/2 + i/n \\ 1/2 - j/n \end{smallmatrix}\right]\!(\hbar/n|\tau)}t_{i,j}\otimes t_{-i,-j}\,,
\]
%{1/2\brack1/2}
in terms of the $\theta$-functions $\theta{a\brack b}(z|\tau) = \sum_{m\in\Z}\exp(\pi i(m+a)^2\tau + 2\pi i(m+a)(z+b))$. It is easy to see that the next-to-leading order contribution to the $R$-matrix in the semi-classical limit is the classical $r$-matrix from equation \eqref{eq:ellrmx}, with $\zeta=1$.

Now we turn our attention to the elliptic $K$-matrices appearing in~\cite{Komori:1997yme}. They depend on the 4 free complex parameters $\lambda_*$, indexed by the pair $r_*,s_*\in\{0,1\}$, together with a further free complex parameter $\Lambda$. This gives 5 complex degrees of freedom, one of which describes the overall scale of the $K$-matrix. We can eliminate this redundancy by requiring
\[ 
	\sum_{r_*,s_*\in\{0,1\}}\lambda_*(-)^{nr_*s_*} = 1\,.
\]
Whilst this family of $K$-matrices is independent of the parameter $\hbar$ appearing in the $R$-matrix, the $K$-matrices generated using gauge theory are formal power series in $\hbar$. The resolution of this apparent inconsistency is that the parameter $\Lambda$ is of order $\hbar$ in the perturbative expansion, and so should be viewed as an element of $\hbar\,\C[[\hbar]]$. To get the correct semi-classical limit it is enough to assume that $\Lambda = \hbar Q$ for $Q\in\C$.

With respect to the standard basis  $\{e_\alpha^{~\beta}\}_{\alpha,\beta\,\in\,\Z_n}$ of ${\rm End}(\C^n)$, we have
\[ 
K(z) = \sum_{r_*,s_*\in\{0,1\}}\lambda_*\sum_{\alpha,\beta\in\Z_n}K^\alpha_{\ \beta}(z;r_*,s_*)\,e_{\alpha}^{~\beta}\,,
\]
where
\[ 
K^\alpha_{~\beta}(z;r_*,s_*) = \frac{\theta\!\left[\begin{smallmatrix} 1/2 - 2\beta/n \\ 1/2 \end{smallmatrix}\right]\!(2Q\hbar-2z|n\tau)\,
\theta\!\left[\begin{smallmatrix} s_*/2 - \alpha/n \\ nr_*/2 \end{smallmatrix}\right]\!(Q\hbar+z|n\tau)\,
\theta\!\left[\begin{smallmatrix}1/2 \\ 1/2\end{smallmatrix}\right]\!(2Q\hbar|n\tau)}{\theta\!\left[\begin{smallmatrix}1/2+(\alpha-\beta)/n \\ 1/2\end{smallmatrix}\right]\!(-2z|n\tau)\,
\theta\!\left[\begin{smallmatrix} s_*/2 - \beta/n \\ nr_*/2 \end{smallmatrix}\right]\!(Q\hbar-z|n\tau)\,
\theta\!\left[\begin{smallmatrix} 1/2 - (\alpha+\beta)/n \\ 1/2 \end{smallmatrix}\right]\!(2Q\hbar|n\tau)}\,.
\]
and we have to chosen to normalize the $K$-matrix conveniently so that its classical limit is
\[ 
	\tau = \lim_{\hbar\to0}K(z) = \sum_{r_*,s_*\in\{0,1\}}\lambda_*\sum_{\alpha,\beta\in\Z_n}(-)^{nr_*s_*}\delta_{\alpha+\beta\equiv0\,(n)}\,e_\alpha^{~\beta} = \sum_{\alpha\in\Z_n}e_\alpha^{~-\alpha}\,.
\]
Conjugating by $\tau$ gives the following automorphism
\[ {\rm conj}_\tau(t_{i,j}) = t_{-i,-j}\,.\]
Viewed as an automorphism of $\mathfrak{sl}_n\C$ via the embedding $\mathfrak{sl}_n\C\subset{\rm End}(\C^n)$, this coincides with one of the automorphisms discussed in subsection \ref{subsec:ellorb}. In particular it corresponds to $\xi=\eta=0$. Given $\tau$ we can form
\[ L(z) = K(z)\tau^{-1} \]
which is somewhat easier to work with. Decomposing $L(z)$ as 
\[ 
	L(z) = \sum_{i,j\in\Z_n}L^{i,j}(z)\,t_{i,j} 
	=  \sum_{r_*,s_*\in\{0,1\}}\lambda_*\sum_{i,j\in\Z_n}L^{i,j}(z;r_*,s_*)\,t_{i,j} 
\]
we have
\begin{multline}
	L^{i,j}(z;r_*,s_*) = \frac{\theta\!\left[\begin{smallmatrix}1/2 \\ 1/2\end{smallmatrix}\right]\!(2Q\hbar|n\tau)}{n\theta\!\left[\begin{smallmatrix} 1/2 - i/n \\ 1/2 \end{smallmatrix}\right]\!(2Q\hbar|n\tau)}\\
	\times\ \sum_{k\in \Z_n}\varepsilon^{kj}\frac{\theta\!\left[\begin{smallmatrix} 1/2 + 2k/n \\ 1/2 \end{smallmatrix}\right]\!(2Q\hbar-2z|n\tau)\,
	\theta\!\left[\begin{smallmatrix} s_*/2 - i/n - k/n \\ nr_*/2 \end{smallmatrix}\right]\!(Q\hbar+z|n\tau)}{\theta\!\left[\begin{smallmatrix}1/2+ i/n + 2k/n \\ 1/2\end{smallmatrix}\right]\!(-2z|n\tau)\,
	\theta\!\left[\begin{smallmatrix} s_*/2 + k/n \\ nr_*/2 \end{smallmatrix}\right]\!(Q\hbar-z|n\tau)}\,.
\end{multline}
To make sense of the expression on the right hand side we determine its quasi-periodicities and pole structure. For the quasi-periodicities we find
\[ 
	L^{i,j}(z+1;r_*,s_*) = \varepsilon^iL^{i,j}(z;r_*,s_*)\,,\qquad L^{i,j}(z+\tau;r_*,s_*) 
	= \varepsilon^je^{2Q\hbar/n}L^{i,j}(z;r_*,s_*)\,.
\]
so it is sufficient to determine the locations and residues of poles in the fundamental domain. In the fundamental domain, $L^{i,j}$ has a simple pole at $z= 0$, $1/2$, $\tau/2$ and $(1+\tau)/2$. However, the residue at this pole differs qualitatively for $n$ even and odd, so we treat the two cases separately.

When $n$ is odd, we find the residues
\[ \begin{aligned}
{\rm Res}_{z=0}\,L^{i,j}(z;r_*,s_*) &= -\frac{1}{2n}\varepsilon^{-2^{-1}ij}\Delta(Q\hbar|n\tau)\,,\\
{\rm Res}_{z=1/2}\,L^{i,j}(z;r_*,s_*) &= -\frac{1}{2n}\varepsilon^{-2^{-1}i(j-1)}(-)^{s_*}\Delta(Q\hbar|n\tau)\,,\\
{\rm Res}_{z=\tau/2}\,L^{i,j}(z;r_*,s_*) &= -\frac{1}{2n}\varepsilon^{-2^{-1}(i-1)j}(-)^{r_*}e^{2\pi i Q\hbar/n}\Delta(Q\hbar|n\tau)\,,\\
{\rm Res}_{z=(1+\tau)/2}\,L^{i,j}(z;r_*,s_*) &= -\frac{1}{2n}\varepsilon^{-2^{-1}(i-1)(j-1)}(-)^{r_*+s_*}e^{2\pi i(Q\hbar + 1/2)/n}\Delta(Q\hbar|n\tau)\,,\\
\end{aligned} \]
where $\Delta(z|\tau) = \theta\!\left[\begin{smallmatrix}1/2 \\ 1/2\end{smallmatrix}\right]\!(z|\tau) \bigg/ \theta\!\left[\begin{smallmatrix} 1/2 \\ 1/2 \end{smallmatrix}\right]'\!(0|\tau)$.
Rescaling the $K$-matrix by
\[ 
-n\frac{\theta\!\left[\begin{smallmatrix} 1/2 \\ 1/2 \end{smallmatrix}\right]'\!(0|n\tau)\,\theta\!\left[\begin{smallmatrix} 1/2 \\ 1/2 \end{smallmatrix}\right]\!(-2Q\hbar/n|\tau)}{\theta\!\left[\begin{smallmatrix}1/2 \\ 1/2\end{smallmatrix}\right]'\!(0|\tau)\,\theta\!\left[\begin{smallmatrix} 1/2 \\ 1/2 \end{smallmatrix}\right]\!(2Q\hbar|n\tau)} = 1 + \mathcal{O}(\hbar)\,,
\]
and introducing parameters ${\hat q}_*$ labelled by $a_*,b_*\in\{0,1\}$ as
\[
 {\hat q}_* = {\hat q}_*(a_*,b_*) = (-)^{na_*b_*}\sum_{r_*,s_*\in\{0,1\}}(-)^{a_*s_*+b_*r_*}\lambda_* \,,
\]
these residues show that, when $n$ is odd, the $L^{i,j}(z)$ corresponding to the rescaled $K$-matrix can be expressed as
\[ \begin{aligned}
L^{i,j}(z) = \sum_{a_*,b_*\in\{0,1\}}\frac{{\hat q}_*}{2}\varepsilon^{-2^{-1}(ij - a_*i - b_*j)}e^{2\pi ib_*Q\hbar/n}\frac{\theta\!\left[\begin{smallmatrix}1/2 \\ 1/2\end{smallmatrix}\right]\!(-2Q\hbar/n|\tau)\,\theta\!\left[\begin{smallmatrix}1/2+i/n \\ 1/2-j/n\end{smallmatrix}\right]\!(z-z_*-2Q\hbar/n|\tau)}{\theta\!\left[\begin{smallmatrix}1/2 + i/n \\ 1/2-j/n \end{smallmatrix}\right]\!(-2Q\hbar/n|\tau)\,\theta\!\left[\begin{smallmatrix}1/2 \\ 1/2 \end{smallmatrix}\right]\!(z-z_*)} 
\end{aligned} \]
where $z_* = (a_*+b_*\tau)/2$. When expressed in terms of the ${\hat q}_*$, our constraint on the $\lambda_*$ becomes
\[ 
	\sum_{a_*,b_*\in\{0,1\}}{\hat q}_* = 1\,.
\]
In this form it is straightforward to take the semi-classical limit. We find that for $i,j\not\equiv0~(n)$
\[ L^{i,j}(z) = - \frac{\hbar}{n}\sum_{a_*,b_*\in\{0,1\}}q_*\varepsilon^{-2^{-1}(ij - a_*i -b_*j)}\frac{\theta\!\left[\begin{smallmatrix}1/2 \\ 1/2\end{smallmatrix}\right]'\!(0|\tau)\ \theta\!\left[\begin{smallmatrix}1/2+i/n \\ 1/2-j/n\end{smallmatrix}\right]\!(z-z_*|\tau)}{\theta\!\left[\begin{smallmatrix}1/2 + i/n \\ 1/2-j/n \end{smallmatrix}\right]\!(0|\tau)\ \theta\!\left[\begin{smallmatrix}1/2 \\ 1/2 \end{smallmatrix}\right]\!(z-z_*|\tau)} + \mathcal{O}(\hbar^2)\,. \]
Up to the addition of a term at order $\hbar$ which is proportional to the identity, we find that
\[
L(z) = {\bf 1} + \frac{\hbar}{n}\sum_{a_*,b_*\in\{0,1\}}q_*\sum_{i,j\in\Z_n}\varepsilon^{-2^{-1}(ij-a_*i-b_*j)}w_{i,j}(z-z_*)t_{i,j} + \mathcal{O}(\hbar^2)\,.
 \]
Here $q_* = -Q{\hat q}_*$, and $w_{i,j}(z-z_*)$ is the unique meromorphic function with a single simple pole at the origin obeying
\[ 
	{\rm Res}_{z=0}\,w_{i,j}(z) = 1\,,\qquad w_{i,j}(z+1) =  \varepsilon^iw_{i,j}(z)\qquad\text{and}\qquad w_{i,j}(z+\tau) = \varepsilon^jw_{i,j}(z)\,.
\]
Note that $Q$ essentially restores the scale of the ${\hat q}_*$, so that the $q_*$ are unconstrained complex numbers. This exactly matches the result derived using gauge theory in section~\ref{subsec:ellKmx}.

\medskip

Now we repeat the computation for $n$ even. The locations and residues of the poles of $L^{i,j}(z;r_*,s_*)$ are listed below
\[ \begin{aligned}
{\rm Res}_{z=0}L^{i,j}(z;r_*,s_*) &= -\frac{1}{n}\varepsilon^{-(i/2)j}\delta_{i\equiv j\equiv 0\,(2)}\Delta(Q\hbar|n\tau)\,,\\
{\rm Res}_{z=1/2}L^{i,j}(z;r_*,s_*) &= -\frac{1}{n}\varepsilon^{-(i/2)(j-1)}(-)^{s_*}\delta_{i\equiv 0\,(2)}\delta_{j\equiv 1\,(2)}\Delta(Q\hbar|n\tau)\,,\\
{\rm Res}_{z=\tau/2}L^{i,j}(z;r_*,s_*) &= -\frac{1}{n}\varepsilon^{-((i-1)/2)j}(-)^{r_*}\delta_{i\equiv 1\,(2)}\delta_{j\equiv 2\,(2)}e^{2\pi i Q\hbar/n}\Delta(Q\hbar|n\tau)\,,\\
{\rm Res}_{z=(1+\tau)/2}L^{i,j}(z;r_*,s_*) &= -\frac{1}{n}\varepsilon^{-((i-1)/2)(j-1)}(-)^{r_*+s_*}e^{2\pi i(Q\hbar + 1/2)/n}\delta_{i\equiv j\equiv 0\,(1)}\Delta(Q\hbar|n\tau)\,,\\
\end{aligned} \]
with $\Delta$ as for $n$ odd. Premultiplying $L(z)$ by
\[ 
	-n\frac{\theta\!\left[\begin{smallmatrix} 1/2 \\ 1/2 \end{smallmatrix}\right]'\!\!(0|n\tau)\,
	\theta\!\left[\begin{smallmatrix} 1/2 \\ 1/2 \end{smallmatrix}\right]\!\!(-2Q\hbar/n|\tau)}
	{\theta\!\left[\begin{smallmatrix}1/2 \\ 1/2\end{smallmatrix}\right]'\!\!(0|\tau)\,
	\theta\!\left[\begin{smallmatrix} 1/2 \\ 1/2 \end{smallmatrix}\right]\!\!(2Q\hbar|n\tau)} 
	= 1 + \mathcal{O}(\hbar)\,,
\]
and defining ${\hat q}_*$ as for odd $n$, we deduce that
\[ 
	L^{2k+a_*,2l+b_*}(z) 
	= {\hat q}_*\varepsilon^{-2kl}e^{2\pi i (Q\hbar + a_*/2)b_*/n}\frac{
	\theta\!\left[\begin{smallmatrix} 1/2 \\ 1/2 \end{smallmatrix}\right]\!\!(-2Q\hbar/n|\tau)\,
	\theta\!\left[\begin{smallmatrix} 1/2 + 2k/n + a_*/n \\ 1/2 - 2l/n - b_*/n \end{smallmatrix}\right]\!\!(z-z_*-2Q\hbar/n|\tau)
	}{
	\theta\!\left[\begin{smallmatrix} 1/2+2k/n+a_*/n \\ 1/2-2l/n-b_*/n \end{smallmatrix}\right]\!\!(-2Q\hbar/n|\tau)\,
	\theta\!\left[\begin{smallmatrix} 1/2 \\ 1/2 \end{smallmatrix}\right]\!\!(z-z_*|\tau)} \,,
\]
where again $a_*,b_*\in\{0,1\}$. Our constraint on the $\lambda_*$ translates into the condition that \smash{${\hat q}_*|_{a_*=b_*=0}=1$}. In this form we can directly take the classical limit to find that, up to the addition of a term at order $\hbar$ which is proportional to the identity,
\[ L(z) = {\bf 1} + \frac{\hbar}{m}\sum_{a_*,b_*\in\{0,1\}}q_*\sideset{}{'}\sum_{k,l\in\Z_m}\varepsilon^{2kl}w_{2k+a_*,2l+b_*}(z-z_*)t_{2k+a_*,2l+b_*} + \mathcal{O}(\hbar^2)\,.\]
Here \smash{$q_* = -Q{\hat q}_*e^{\pi i a_*b_*/n}$}, so that the $q_*$ are unconstrained complex parameters. This exactly matches the result derived in section~\ref{subsec:ellKmx}.

\section{Determination of allowed bundle morphisms in the elliptic case}
\label{app:EllipticBundleMorphisms}

In this appendix we will show that, in the elliptic case, the only allowed bundle morphisms $\sigma_Z:{\rm Ad}\,P\to{\rm Ad}\,P$ we can use to construct the orbifold gauge theory are those that were considered in section~\ref{subsec:ellorb}.
In particular, unlike the trigonometric case, holomorphy and compactness ensure that there are no non-constant morphisms.

We first describe ${\rm Ad}\,P$ more explicitly. We begin by noticing that $M=\Sigma\times E_\tau$ is the quotient of the covering space $M'=\Sigma\times\C$ by $(p,z)\sim (p,z-a-b\tau)$ for $a,b\in\Z$. We can pull back ${\rm Ad}\,P$  to $M'$ whereupon, since $\C$ is contractible (and we assume ${\rm Ad}\,P$ is trivial over $\Sigma$), the resulting bundle is isomorphic to the trivial bundle $M'\times\mathfrak{sl}_n\C$. In fact we can choose a trivialization such that the pullback of the vacuum is the partial connection $\d_{\Sigma}+{\bar\partial}_C$. Then ${\rm Ad}\,P$ is the quotient of the covering space $M'\times\mathfrak{sl}_n\C$ by $((p,z),X)\sim((p,z-a-b\tau),{\rm conj}_{A^{\zeta a}B^b}(X))$.

Next we consider the bundle morphism that covers the $\Z_2$ action on the underlying orbifold. The map $Z$ lifts to a map $Z'$ on $M'$ again given by $Z':(x,y,z,\bar z)\mapsto (-x,y,-z,-\bar z)$. Similarly $\sigma_Z$ lifts to a map 
\[ 
	\sigma_Z': M'\times\mathfrak{sl}_n\C\to M'\times\mathfrak{sl}_n\C\,,
\]
where explicitly
\[
	\sigma_Z': (w,X)\mapsto (Z'(w),\sigma_Z'(w;X))\,.
\]
Since $\sigma_Z$ must preserve the $G$-structure of ${\rm Ad}\,P$, the map $X\mapsto\sigma_Z'(w;X)$ must be an automorphism of $\mathfrak{sl}_n\C$. To preserve the vacuum it must be constant on $\Sigma$ and vary holomorphically over $\C$; we indicate this by writing $\sigma_Z'(w;X)=\sigma_Z'(z;X)$. For $\sigma'_Z$ to descendconsistently to a map on the bundle ${\rm Ad}\,P$ over the elliptic curve, over $\C$ we must have
\be 
\label{eq:sigmaphase} 
	\sigma_Z'(z;X) = 
	{\rm conj}_{A^{\zeta a}B^{b}}(\sigma_Z'(z+a+b\tau;{\rm conj}_{A^{\zeta a}B^b}X))
\ee
for all $z\in\C$ and all $X\in\mathfrak{sl}_n\C$, and where $a,b\in\Z$. %Recall that $\sigma'_Z$ is involutive, so
%\[ \sigma'_Z(z;\sigma'_Z(-z;X)) = X \]
%for all $X\in\mathfrak{sl}_n\C$.
Since $X\mapsto\sigma'_Z(w;X)$ is a Lie algebra automorphism it is linear and so, in terms of the basis $\{t_{i,j}\}_{(i,j)\in\mathcal{I}_n}$ of $\mathfrak{sl}_n\C$, we may write
\[ 
	\sigma'_Z(z;t_{i,j}) = \sum_{k,l}\sigma^{k,l}_{~~i,j}(z)\,t_{k,l} 
\]
for $\sigma^{k,l}_{~~i,j}(z)$ holomorphic functions of $z$. When expressed in this basis, equation~\eqref{eq:sigmaphase} tells us that
\[ 
	\sigma^{k,l}_{~~i,j}(z+a+b\tau) \,\varepsilon^{a(k+i)+b(j+l)}  
	= \sigma^{k,l}_{~~i,j}(z)\,.
\]
We learn that the $\sigma^{k,l}_{~~i,j}(z)$ are bounded, and since they are holomorphic they must also be constant. This shows that $\sigma_Z$ acts as a constant involutive automorphism on the fibres of ${\rm Ad}\,P$. A constant function is clearly periodic, hence
\[ 
	\sigma^{k,l}_{~~i,j} = \delta_{i+k\equiv0\,(n)}\,\delta_{j+l\equiv0\,(n)}\,\mu_{i,j} 
\]
for complex numbers $\mu_{i,j}\in\C$.

Now we require that this defines an automorphism of $\mathfrak{sl}_n\C$. For this map to have trivial  kernel the $\mu_{i,j}$ must all be non-vanishing. In the basis  $\{t_{i,j}\}_{(i,j)\in\mathcal{I}_n}$ the Lie bracket is given by
\[ 
	[t_{i,j},t_{k,l}] 
	= (\varepsilon^{-\zeta^{-1}jk} - \varepsilon^{-\zeta^{-1}il})\,t_{i+k,j+l}\,
\]
and we learn that
\[ 
	(\mu_{i+k,j+l} - \mu_{i,j}\mu_{k,l})
	(\varepsilon^{-\zeta^{-1}jk} - \varepsilon^{-\zeta^{-1}il}) = 0\,.
\]
Thus $\mu_{i+k,j+l} = \mu_{i,j}\mu_{k,l}$ whenever $jk\not\equiv il~(n)$. This allows to build up $\mu_{i,j}$ recursively from $\mu_{1,0}$ and $\mu_{0,1}$, and we 
find that the recursion relation is satisfied by $\mu_{i,j} = (\mu_{1,0})^i(\mu_{0,1})^j$ for all $(i,j)\in\mathcal{I}_n$. 

Finally, notice that $\mu_{0,1}^{n+1}\mu_{1,0} = \mu_{0,1}\mu_{1,0}$ so that $\mu_{1,0}^n=1$, or in other words $\mu_{1,0}$ is an $n^\text{th}$ root of unity. In particular $\mu_{1,0}=\varepsilon^\xi$ for some $\xi\in\Z_n$, and similarly $\mu_{0,1}=\varepsilon^\eta$ for some $\eta\in\Z_n$. This allows us to conclude that
\[ 
	\sigma_Z': \big((x,y,z,\bar z),t_{i,j}\big)\mapsto \big((-x,y,-z,-\bar z),\varepsilon^{i\xi+j\eta}t_{-i,-j}\big)\,. 
\]
This is the  involutive automorphism that was used in equation~\eqref{eqn:ellipticsigma}.

\section{Classification of permissible involutive automorphisms of $\fg = \fg_0\oplus {\widetilde \fh}$}
\label{app:3}
In this appendix we classify involutive automorphisms of the Lie algebra \smash{$\fg = \fg_0\oplus{\widetilde \fh}$} which swap the maximal isotropic subalgebras $\fg_+$ and $\fg_-$, and preserve an invariant bilinear on $\fg$. Here \smash{$\fg_0$} is a complex, simple Lie algebra, and \smash{${\widetilde \fh}$} a second copy of its Cartan subalgebra. The subalgebras $\fg_+$ and $\fg_-$ are defined by
\[
	\fg_+ =\mathfrak{n}_+\oplus\{(H,i\widetilde H)|H\in\fh\}\,,\qquad
	\fg_- =\mathfrak{n}_-\oplus\{(H,iM({\widetilde H}))|H\in\fh\}\,.
\]
for $M\in {\rm End}({\widetilde \fh})$. Recall that for these subalgebras to be disjoint $M$ must not have a +1 eigenvalue. We seek an involutive automorphism $\sigma$ of $\fg$, such that
\[ 
	\sigma(\fg_+) = \fg_-\,.
\]
This automorphism must preserve the invariant bilinear 
\[ 
	\langle~,~\rangle_\fg 
	= \langle~,~\rangle_{\fg_0} + \langle~,~\rangle_{\widetilde \fh}
\]
on $\fg$, where \smash{$\langle~,~\rangle_{\fg_0}$} is proportional to the Killing form on $\fg_0$, and \smash{$\langle~,~\rangle_{\widetilde \fh}$} is its restriction to the Cartan. Bulk integrability requires that $\fg_-$ is isotropic, which forces $M$ to be orthogonal with respect to \smash{$\langle~,~\rangle_{\widetilde \fh}$}.

\medskip

We begin by decomposing our automorphism with respect to the direct sum $\fg = \fg_0\oplus{\widetilde \fh}$, writing
\[ 
	\sigma = \begin{pmatrix}
		\chi & \phi \\
		\psi & \omega
	\end{pmatrix}\,.
\]
Since the centre of $\fg$ is $\widetilde{\fh}$ and since any Lie algebra automorphism preserves the centre, we must have $\sigma({\widetilde \fh})\subseteq{\widetilde\fh}$. Hence $\phi=0$. For $\sigma$ to be involutive we actually require $\sigma({\widetilde\fh})={\widetilde\fh}$ with $\omega^2={\bf 1}_{\widetilde\fh}$.

The requirement that $\sigma$  preserves our chosen bilinear implies that $\psi=0$ also. This is because $\langle \widetilde{H},X\rangle_\fg =0$ for any $X\in\fg_0$ and any $\widetilde{H} \in\widetilde\fh$, so preservation of $\langle~,~\rangle_\fg$ in particular implies
\[ 
	0=\langle\sigma(\widetilde H),\sigma(X)\rangle_\fg 
	= \langle\sigma(\widetilde H),\chi(X) + \psi(X)\rangle_\fg 
	= \langle\sigma(\widetilde H),\psi(X)\rangle_{\widetilde \fh} 
\]
where the third equality follows since $\sigma$ preserves $\widetilde\fh$. The bilinear \smash{$\langle~,~\rangle_{\widetilde h}$} is non-degenerate, so we have $\psi=0$ as claimed. 

$\sigma$ being involutive also implies $\chi^2={\bf 1}_{\fg_0}$, so that $\chi$ is an involutive automorphism of $\fg_0$. Indeed, we can interpret $\chi$ as the action of $\sigma$ on the quotient \smash{$\fg/{\widetilde\fh}\cong\fg_0$}. When interpreted on this quotient, the statement that $\sigma$ swaps the subalgebras $\fg_+$ and $\fg_-$ implies that $\chi$ swaps the Borel subalgebras
\[ 
	\mathfrak{b}_+ = \mathfrak{n}_+\oplus \fh\,,\qquad \mathfrak{b}_- 
	= \mathfrak{n}_-\oplus \fh\,.
\]
We require that $\chi(\mathfrak{b}_+)=\mathfrak{b}_-$, and since $\chi$ is involutive this implies that $\chi(\mathfrak{b}_-)=\mathfrak{b}_+$. Hence
\[ 
	\chi(\fh) \subset \left(\mathfrak{b}_+\cap\mathfrak{b}_-\right)=\fh 
\]
and we lean that $\chi$ fixes $\fh$.
% We abuse notation by writing $\chi$ for the restriction of $\chi$ to $\fh$.
%Note that any automorphism of a simple Lie algebra leaves the Killing form invariant, so we get no constraints on $\chi$ from the requirement that $\sigma$ preserves our invariant bilinear.
Since $\chi$ preserves the Cartan, it must map root spaces to other root spaces. Furthermore, since $\chi$ swaps the Borel subalgebras $\mathfrak{b}_\pm$ it must swap positive roots and negative roots. This restricts the action of $\chi$ on the roots to be the composition of multiplication by $-1$ with an involutive automorphism of the Dynkin diagram, which we will denote by $\gamma\in{\rm Sym}(\Delta)$.

Now let $s$ and $\pi_\gamma$ be the involutive automorphisms
\[ 
\begin{aligned}
	s:~&h_\mu\mapsto -h_\mu\,,\qquad
	e_\mu\mapsto f_\mu\,,\qquad
	f_\mu\mapsto e_\mu \\
	\pi_\gamma:~&h_\mu\mapsto h_{\gamma(\mu)}\,,\qquad
	e_\mu\mapsto e_{\gamma(\mu)}\,,\qquad
	f_\mu\mapsto f_{\gamma(\mu)}
 \end{aligned} 
\]
in terms of the Chevalley basis  $\{f_\mu,e_\mu\}_{\mu\in\Phi_+}\cup\{h_\mu\}_{\mu\in\Delta}$ of $\fg_0$. When $\mu\in\Delta$, we find that the composition $\pi_\gamma\circ s\circ\chi$ fixes $\fh$ pointwise. (The fact that the above formulas can be uniquely extended to involutive automorphisms of $\fg_0$ is a standard result in elementary Lie algebra theory; see {\it e.g.} \cite{Humphreys:1972}). Any automorphism fixing $\fh$ pointwise must fix the root spaces, and so the only freedom remaining in $\pi_\gamma\circ s\circ\chi$ is to map \smash{$e_\mu\mapsto \Lambda_\mu e_\mu$ } for $\mu\in\Delta$, where $\Lambda_\mu\in\C^*$. Since $[e_\mu,f_\mu]=h_\mu$ we must then have $f_\mu\mapsto\Lambda_\mu^{-1} f_\mu$ also. This can be neatly summarised by writing
\[ 
	\chi = s\circ\pi_\gamma\circ\exp({\rm ad}_\lambda)\,,
\]
where $\lambda = \lambda^\mu h_\mu\in\fh$ is given by $\exp(\sum_\nu A_{\mu\nu}\lambda^\nu) = \Lambda_\mu$ for $A$ the Cartan matrix of $\fg_0$. For this $\chi$ to be involutive we require that $\lambda$ lies in the +1 eigenspace of $\pi_\gamma|_\fh$.

\medskip

Whether we get a solution of the `soliton preserving' or `soliton reversing' bYBE depends on whether $\chi$ is an inner or outer automorphism. If $\fg_0$ is not of type $A_n$, $D_n$, or $E_6$ then there are no outer automorphisms, and so $\chi$ is necessarily inner. In the remaining cases, certainly $\exp({\rm ad}_\lambda)$ is inner, any non-trivial $\pi_\gamma$ is outer, and $s$ can be inner or outer depending on $\fg_0$. In fact, $s$ is always outer when $\fg_0=A_n$ with $n\geq2$,  when $\fg_0=D_n$ with $n\geq4$, $s$ is inner for even $n$ and outer for odd $n$ ($s$ swaps the two spin representations), and finally $s$ is outer for $E_6$.

\medskip

Having constrained $\chi$ it remains to determine $\omega$. For $\sigma$ to preserve $\langle~,~\rangle_\fg$ we must have $\omega\in O({\widetilde \fh})$, while for $\sigma$ to be involutive we must have $\omega^2={\bf 1}_{\widetilde\fh}$. We still need to impose $\sigma(\fg_+) = \fg_-$. Consider $(H,i{\widetilde H})\in\fg_+$ for some $H\in\fh$. Acting with $\sigma$ we learn that
\[ 
	\sigma((H,i{\widetilde H})) 
	= (\chi(H),i\omega({\widetilde H})) = (H',iM({\widetilde H}'))\in\fg_-
	\]
for some $H'=\chi(H)\in\fh$, so we must have $\omega = M\circ\chi$. (Here we've restricted $\omega$ to the Cartan and interpreted it as an endomorphism of $\widetilde\fh$.) Since we've already imposed the condition that $\sigma$ be involutive, this is sufficient to ensure that $\sigma(\fg_-)=\fg_+$. The condition $\omega^2=1$ imposes the constraint
\[ 
	(M\circ\chi)^2 = {\bf 1}_{\widetilde\fh}
\]
on $M$. Since $M\in O({\widetilde\fh})$ and $\chi|_\fh\in O(\fh)$, it follows immediately that $\omega=M\circ\chi\in O({\widetilde\fh})$.

In the case that we choose $\gamma$ to be the identity, we have $\chi|_{\fh} = -{\bf 1}_\fh$, and $M$ is constrained by $M^2={\bf 1}$. Any involutive $M$ is diagonalizable with $+1$ and $-1$ eigenspaces, however $M$ cannot have a $+1$ eigenvalue. We deduce that $M=-{\bf 1}_{\widetilde\fh}$, and $\omega={\bf 1}_{\widetilde\fh}$. On the other hand, if we choose $\gamma$ to be a non-trivial automorphism of the Dynkin diagram then there exist non-trivial choices for $M$.

\section{An aside on loop algebras}
\label{app:loopalg}

In order to generate candidates for $z$-dependent automorphisms of $\fg$, we turn our attention to the loop group $L\fg = \fg[u,u^{-1}]$ of finite Laurent series in the formal parameter $u$. The motivation for this is that certain families of automorphisms of the loop algebra are given by conjugation by a $z$-dependent automorphism of $\fg$, and automorphisms of the loop algebras can be constructed using standard results from the theory of affine algebras. Note that the loop algebra has the structure of a Manin triple with respect to the natural symmetric bilinear
\[ 
	(a(u),b(u))_{L\fg} = \oint\frac{\diff u}{u}(a(u),b(u))_{\fg}\,.
\]
The isotropic subalgebras are given by
\[ 
	(L\fg)_- = \fg[u]\oplus\fg_-\qquad\text{and}\qquad
	 (L\fg)_+ = \fg[u^{-1}]\oplus\fg_+\,.
\]

We would like to find automorphisms of $L\fg$ which can be realised as $z$-dependent automorphisms of the algebra $\fg$. We shall again denote such automorphisms by $\sigma$.  The loop algebra admits the involution $\imath :u\mapsto u^{-1}$, which allows us to restate the condition $\sigma(-u)\circ\sigma(u) = {\rm id.}$ as requiring that $\theta = \sigma\circ \imath$ is involutive. The condition that $\sigma$ preserves the boundary conditions at $z=\pm\infty$, together with the fact that it should map non-singular configurations to non-singular configurations implies that $\theta$ should swap $(L\fg)_+$ and $(L\fg)_-$.

To find such automorphisms, we first quotient out the centre of this algebra, which is simply $L\widetilde\fh$. $\theta$ descends to the quotient, and determines an involutive automorphism of $L\fg_0$ swapping the subalgebras $\mathfrak{b}_- = \mathfrak{h}\oplus \mathfrak{n}_-\oplus\fg_0[u^{-1}]$ and $\mathfrak{b}_+ = \mathfrak{h}\oplus \mathfrak{n}_+\oplus\fg_0[u]$.
We can lift this automorphism to the affine algebra $\widehat{\fg}_0'$, and we learn that $\theta$ must exchange two of its Borel subalgebras. (The Borel subalgebras are the $\mathfrak{b}_\pm$ defined above with $\fh$ extended to include the central element $c$.) Such automorphisms can easily be classified, following arguments presented in~\cite{Kac:1992aut} and in the appendix of~\cite{Kolb:2014qu}. One finds that all automorphisms swapping these subalgebras are of the form
\[ 
	\theta = {\rm Ad}(\Lambda) \circ \Gamma \circ \omega \,,
\]
where
\[ 
\omega: (e_\mu,f_\mu,h_\mu) \mapsto (f_\mu,e_\mu,-h_\mu)
\qquad\text{for $\mu\in\widehat\Delta$} 
\]
is the Chevalley involution, where $\Gamma$ extends the action of a permutation of the Dynkin diagram of $\widehat{\fg}_0'$, $\gamma$, to all of $\widehat{\fg}_0'$ by
\[ 
\Gamma: (e_\mu,f_\mu,h_\mu) \mapsto 	
(e_{\gamma(\mu)},f_{\gamma(\mu)},h_{\gamma(\mu)})
\qquad\text{for  $\mu\in\widehat\Delta$} \,,
\]
and finally where
\[ 
	{\rm Ad}(\Lambda):(e_\mu,f_\mu,h_\mu)
	\mapsto:(\Lambda(\alpha_\mu) e_\mu,\Lambda(\alpha_\mu)^{-1} f_\mu, h_\mu)
\]
for some $\gamma$-invariant map $\Lambda:\widehat\Delta\mapsto \C^*$. Here $\widehat\Delta$ denotes the set of simple roots of the affine algebra $\widehat{\fg}_0'$, and $\{e_\mu,f_\mu,h_\mu\}$ are its Chevalley generators. Given such a $\theta$ we can restrict it to the loop algebra $L\fg_0$, and then extend it to $L\widetilde\fh$ by defining $\theta|_{\widetilde\fh} = M\circ\theta|_\fh$. Indeed this is the only way of extending it consistent with the fact that $\theta$ must swap $(L\fg)_-$ and $(L\fg)_+$. Finally by taking the composition $\theta\circ \imath$ we hope to recover a $z$-dependent automorphism of $\fg$ of the required form. Unfortunately not all maps generated in this way are $z$-dependent automorphisms of $\fg$. (A simple example of an automorphisms of $L\fg$ which is not a $z$-dependent automorphism of $\fg$ is the map $z\mapsto \lambda z$ for $z\in\C$.) Fortunately by judiciously choosing the parameters in the function $\Lambda$ we can generate automorphisms of the desired form.

This is how the example automorphisms in the following section were generated. It would be interesting to explore how to realise $K$-matrices associated to more general involutive automorphisms of the second kind, as described in~\cite{Kolb:2014qu}.

\vspace{1cm}

\noindent {\large{\bf Acknowledgements}}

This work has been partially supported by STFC consolidated grant ST/P000681/1. The work of RB is supported by EPSRC studentship EP/N509620/1.

\vspace{1cm}

\bibliographystyle{JHEP}
\bibliography{references}

\end{document}